\documentclass[opre,nonblindrev]{informs3-rad}

\TheoremsNumberedThrough     
\ECRepeatTheorems

\EquationsNumberedThrough    

\MANUSCRIPTNO{OPRE-2021-01-011} 

\usepackage[parfill]{parskip}
\usepackage{mathtools}
\usepackage{mathrsfs} 
\usepackage{natbib}
 \bibpunct[, ]{(}{)}{,}{a}{}{,}%
 \def\bibfont{\small}%
 \def\bibsep{\smallskipamount}%
\usepackage{ifthen}
\usepackage{algpseudocode,algorithm,algorithmicx}

\usepackage[breakable, skins]{tcolorbox}
\DeclareRobustCommand{\mybox}[2][gray!20]{%
\begin{tcolorbox}[   
        left=0pt,
        right=0pt,
        top=0pt,
        bottom=0pt,
        colback=#1,
        colframe=#1,
        width=\dimexpr\textwidth\relax, 
        enlarge left by=0mm,
        boxsep=5pt,
        arc=5pt,outer arc=5pt,
        ]
        #2
\end{tcolorbox}
}

\usepackage{csquotes}
\makeatletter
\renewenvironment*{displayquote}
  {\begingroup\setlength{\leftmargini}{0.2cm}\csq@getcargs{\csq@bdquote{}{}}}
  {\csq@edquote\endgroup}
\makeatother

\usepackage{dsfont}
\definecolor{cornellred}{rgb}{0.7, 0.11, 0.11}
\definecolor{dgreen}{rgb}{0.0, 0.5, 0.0}
\definecolor{ballblue}{rgb}{0.13, 0.67, 0.8}
\definecolor{royalblue(web)}{rgb}{0.25, 0.41, 0.88}
\definecolor{bleudefrance}{rgb}{0.19, 0.55, 0.91}
\definecolor{royalazure}{rgb}{0.0, 0.22, 0.66}
\usepackage{hyperref}
\hypersetup{
    colorlinks = true,
    linkcolor=cornellred,
    citecolor=royalazure,
    linkbordercolor = {white}
}
\usepackage{cleveref}
\usepackage[shortlabels]{enumitem}
\usepackage{mathrsfs}

\newcommand{\expect}[1]{\ensuremath{{\mathbb E}[#1]}}

\DeclareMathOperator{\patience}{\theta}

\DeclarePairedDelimiterX{\set}[1]\{\}{#1}
\let\Pr\relax
\DeclarePairedDelimiterXPP{\Pr}[1]{\mathbb{P}}[]{}{#1}
\DeclarePairedDelimiterXPP{\Ex}[1]{\mathbb{E}}[]{}{#1}

\newcommand{\setoftypes}{\mathcal{U}}
\newcommand{\tpi}{\pi}
\newcommand{\Rev}{\mathsf{Rev}}

\newcommand{\revcolor}[1]{{#1}}

\begin{document}
\RUNAUTHOR{Asadpour et al.}

\RUNTITLE{Sequential Submodular Maximization and Applications to Ranking an Assortment of Products }

\TITLE{Sequential Submodular Maximization and \\Applications to Ranking an Assortment of Products}

\ARTICLEAUTHORS{%
\AUTHOR{Arash Asadpour}
\AFF{Zicklin School of Business, City University of New York, \EMAIL{arash.asadpourrahimabadi@baruch.cuny.edu}.}
\AUTHOR{Rad Niazadeh}
\AFF{Chicago Booth School of Business, University of Chicago, \EMAIL{rad.niazadeh@chicagobooth.edu}.}
\AUTHOR{Amin Saberi}
\AFF{Management Science and Engineering, Stanford University, \EMAIL{saberi@stanford.edu}.}
\AUTHOR{Ali Shameli}
\AFF{Meta, Core Data Science, \EMAIL{alishameli@gmail.com }.}

}

\ABSTRACT{

We study a submodular maximization problem motivated by applications in online retail. A platform displays a list of products to a user in response to a search query. The user inspects the first $k$ items in the list for a $k$ chosen at random from a given distribution, and decides whether to purchase an item from that set based on a choice model. The goal of the platform is to maximize the engagement of the shopper defined as the probability of purchase. This problem gives rise to a less-studied variation of submodular maximization in which we are asked to choose an \emph{ordering} of a set of elements to maximize a linear combination of different submodular functions. 

First, using a reduction to maximizing submodular functions over matroids, we give an optimal  $\left(1-1/e\right)$-approximation for this problem. We then consider a variant in which the platform cares not only about user engagement, but also about diversification across various groups of users, that is, guaranteeing a certain probability of purchase in each group.  We characterize the polytope of feasible solutions and give a bi-criteria $((1-1/e)^2,(1-1/e)^2)$-approximation for this problem by rounding an approximate solution of a linear programming relaxation. For rounding, we rely on our reduction and the particular rounding techniques for matroid polytopes. For the special case in which underlying submodular functions are coverage functions -- which is practically relevant in online retail --  we propose an alternative LP relaxation and a simpler randomized rounding for the problem. This approach yields to an optimal bi-criteria $(1-1/e,1-1/e)$-approximation algorithm for the special case of the problem with coverage functions.

}



\maketitle

 \DoubleSpacedXI
\newenvironment{myfont}{\fontfamily{pag}\selectfont}{\par}

 \vspace{-5mm}
\section{Introduction}
\label{sec:intro}

\revcolor{The advents of online retailing and advertising have created new opportunities for  online platforms to incorporate algorithmic techniques to improve shoppers' experience and drive user engagement, which in return can help with the long-term growth of these platforms.  One such opportunity is optimizing the \emph{combinatorial configuration} of displayed products or ads in response to a search query. This configuration choice can be about selecting a subset of products with certain properties, e.g., of a certain size, or about selecting a ranked list of the products, where higher items in the list are presumably more likely to be clicked by a user. We focus on the case where the seller aims to pick a configuration that maximizes the probability of click (or purchase).\footnote{\revcolor{This is in contrast to product-weighted objectives such as revenue which are more common in the revenue management literature; see for example the line of work on cardinality constrained assortment optimization such as~\citealt{rusmevichientong2010dynamic,gallego2014constrained}, or \citealt{desir2015capacity}).}}}

\revcolor{As an example of the above problem, consider a seller who presents a subset of products of size $k$ to shoppers who may purchase one of the items. In this model, parameter $k$ plays the role of the shopping window's size. As is common in the literature, shoppers' choices are assumed to be probabilistic where shoppers' choice functions determine the click (or purchase) probabilities given any assortment of products. Under the standard substitution assumption~\citep{KFV-08}, the probability of purchase from any assortment is monotone as a function of this set and has decreasing marginal values. As a result, picking an assortment of a given size in order to maximize this purchase probability is an instance of this classic problem: given a monotone submodular function $f:2^{[n]} \rightarrow \mathbb{R}^+$, find a subset $S \subseteq [n]$ of cardinality $k$ that maximizes $f(S)$. This problem has a wide range of applications in economics, operations, and combinatorial optimization~\citep{kempe2003maximizing,krause2014submodular}. The seminal result of \cite{nemhauser1978analysis} shows the greedy algorithm achieves a $(1-1/e)$-approximation for the problem. Moreover, \cite{feige} proves no polynomial-time algorithm can obtain a better approximation ratio unless $\textrm{P = NP}$. }


Motivated by product ranking applications in online retail -- which can range from displaying grocery items on Amazon to displaying a list of rooms on Airbnb -- we study a generalization of the above problem.
In these applications, the displayed products are not only a subset selected by the platform, but are also ranked, typically in a vertical list. 
The  shopper scrolls down the list -- to a point depending on his or her \emph{patience level} -- and potentially purchases one of those products. Again, the probability of purchase depends on his or her choice function, which can be different for shoppers with different patience levels. The  goal of the platform is to pick a ranking to maximize the probability of purchase given the (joint) distribution over the patience levels and the choice functions. This objective is usually referred to as \emph{user engagement}.\footnote{\revcolor{Although our focus is on the probability of purchase, in practice, the platform may take into consideration several other objectives including the revenue or a combination of revenue and social welfare. We show the extension of one of our results for the revenue objective in \Cref{subsec:revenue} and discuss future directions related to these objectives in \Cref{sec:conclusion}.}} This problem is an instance of the following general problem:
\vspace{2mm}
\mybox{
\begin{displayquote}
\begin{problem}[\emph{Sequential Submodular Maximization}]
\label{pbm:seq}
Given monotone submodular functions $f_1, \ldots,  f_n:2^{[n]} \rightarrow \mathbb{R}^+$ and non-negative coefficients $\lambda_1, \ldots,  \lambda_n$, find a randomized permutation $\pi = (\pi_1, \ldots, \pi_n)$ over elements in $[n]$ (i.e., a distribution $D_\pi$ over permutations) in order to 
$$ \underset{D_\pi}{\textrm{maximize}}~\mathbb{E}_{\pi\sim D_\pi}\left[\sum_{i = 1}^n \lambda_i f_i(\{\pi_1, \ldots, \pi_i\})\right].$$
\end{problem}
\end{displayquote}}
\vspace{-2mm}
A few explanations are in order. First, $\lambda_i$ can be thought of as the proportion of users with patience level $i$ and $f_i$ can be thought of as the purchase probability function of users with patience level $i$. The manner in which $f_i$'s are derived from the choice functions and justification for the submodularity assumption are discussed in Section~\ref{subsec:application}. Second, the above optimization problem is not over subsets of elements but over (possibly randomized) sequences. For this reason, we refer to it as a sequential submodular maximization problem. \revcolor{Third, we allow randomized solutions, but the optimal solution of the above program is always a deterministic permutation $\pi^*$.} Finally, note this problem generalizes monotone submodular maximization subject to a cardinality constraint, which is basically the special case when $\lambda_k=1$ for a given $k$, and $\lambda_i=0$ for all $i\neq k$.

\vspace{-2mm}
The platform may also want to take other considerations into account in order to improve the quality of product ranking. One such consideration is to produce a diversified portfolio of user engagements. This  diversification can be utilized to avoid ignoring or marginalizing a group of users.  
More specifically, in addition to maximizing overall user engagement, the platform aims to achieve \emph{group fairness}, that is, guaranteeing a minimum level of engagement for various (possibly overlapping) groups of users. This problem is formulated as the following:

 \mybox{
\begin{displayquote}
\begin{problem}[\emph{Sequential Submodular Maximization with Group Constraints}]
\label{pbm:grp} Given monotone submodular functions $f_1, \ldots, f_n:2^{[n]} \rightarrow \mathbb{R}^+$ and $f^l_1, \ldots, f^l_n:2^{[n]}\rightarrow \mathbb{R}^+$ for each group $l\in[L]$, non-negative coefficients $\lambda_1, \ldots, \lambda_n$ and $\lambda^l_1, \ldots, \lambda^l_n$  for each group $l\in[L]$, and non-negative thresholds $\{T_l\}_{l\in[L]}$, find a randomized permutation $\pi = (\pi_1, \ldots, \pi_n)$ over elements in $[n]$ (i.e., a distribution $D_\pi$ over permutations) in order to
\begin{align*}
\begin{array}{lll}
     \underset{D_\pi}{\textrm{maximize}}&\mathbb{E}_{\pi\sim D_\pi}\left[\displaystyle\sum_{i = 1}^n \lambda_i f_i(\{\pi_1, \ldots, \pi_i\})\right]&~~~~s.t.\\[20pt]
     &\mathbb{E}_{\pi\sim D_\pi}\left[\displaystyle\sum_{i = 1}^n \lambda^l_i f^l_i(\{\pi_1, \ldots, \pi_i\})\right]\geq T_l&~~~~\forall l\in[L].
\end{array}
\end{align*}
\end{problem}
\end{displayquote}}
In this new formulation, $f_i^{l}$ should be thought of as the purchase probability function of users with patience level $i$ in group $l$, and $f_i$ as the aggregate purchase probability function of users with patience level $i$. Similarly, $\lambda_i^l$ should be thought of as the proportion of users with patience level $i$ in group $l$, and $\lambda_i$ as the aggregate proportion of users with patience level $i$. Again, the formal definition of groups (based on the \emph{user types}), the manner in which  $f_i^{l}$ and $f_i$ are derived form the choice functions, and how they are mathematically connected to each other are discussed in \Cref{subsec:application}. \revcolor{Finally, we emphasize that
(i) the optimal solution might be randomized, and (ii) the problem might be infeasible, but becomes
feasible after decreasing some of $T_l$'s; both of these new features are in contrast to Problem 1. (See the discussion at the beginning of \Cref{revenue}.)}

\subsection{Overview of Technical Contributions}


For the problem of maximizing a sequential submodular function (Problem~\ref{pbm:seq}), we present an optimal $(1-1/e)$-approximation algorithm in \Cref{engagement}. Our algorithm is based on a reduction to submodular maximization subject to a (laminar) matroid constraint. The reduction relies on two  ideas. The first  is lifting the problem to a larger space where every element is copied $n$ times, and then defining a certain submodular function and laminar matroid in this larger space that captures maximization over permutations in the original problem. The second is a post-processing that, given a feasible base of the laminar matroid  returns a permutation by only increasing the objective function. For the reduced problem, we use the known approximation algorithm for monotone submodular functions subject to matroids~\citep{calinescu2011maximizing}. 
This result improves the theoretical results of \cite{ferreira} by improving the approximation factor of the offline optimization problem studied in this paper, which is indeed a special case of (unconstrained) sequential submodular maximization.


The approximation algorithm in \cite{calinescu2011maximizing} starts by approximately optimizing a continuous relaxation of the underlying submodular function -- known as the multi-linear extension~\citep{vondrak2011submodular} --  over the matroid polytope. This step is done by running the \emph{continuous greedy algorithm}, a  variant of Franke-Wolfe algorithm used in convex and non-convex optimization~\citep{frank1956algorithm}. The resulting approximately optimal fractional point can then be rounded to a base of the matroid by dependent randomized rounding algorithms such as \emph{swap rounding}~\citep{chekuri2010swap}.\footnote{\revcolor{We highlight that while continuous greedy is a first order method and relatively fast, the gradient access requires sampling the underlying submodular function at enough number of points and can be demanding in terms of sample complexity; standard tricks in first order optimization, e.g., using stochastic gradient instead of gradient and variance reduction~\citep{mokhtari2020stochastic}, can help with running this algorithm on larger practical instances. This is in particular helpful in our application, as one can easily obtain the required stochastic gradients by sampling sets and customer choices over these sets.}}

Next we switch to the problem of maximizing a sequential submodular function subject to group constraints (Problem~\ref{pbm:grp}). 
Relaxing the exact satisfaction of the constraints to an approximation is necessary because satisfying each group constraint on its own is an NP-hard problem. As a result, we settle for \emph{bi-criteria} approximation algorithms, which return randomized permutations that (i) obtain an approximately optimal objective value and (ii) approximately satisfy the group constraints, both in expectation. We highlight that randomized solutions are in particular well-motivated in our applications domain (see \Cref{subsec:application}), as the platform aims to obtain high overall user engagement and satisfy the group constraints on average over time -- which are indeed achievable by the law of large numbers.

To design bi-criteria approximation algorithms for Problem~\ref{pbm:grp}, it is tempting to use our previous reduction, but this time with different submodular functions in the objective and group constraints. Interestingly, the swap rounding algorithm does \emph{not} depend on the choice of the underlying submodular function. So, an approximately optimal and approximately feasible fractional point for the continuous relaxation of  Problem~\ref{pbm:grp} can be rounded using swap rounding (modulo a simple post-processing due to our previous reduction); nevertheless, how to extend the continuous greedy algorithm to handle non-convex constraints or multiple submodular functions in the continuous relaxation of Problem~\ref{pbm:grp} after applying our reduction is unclear. To the best of our knowledge, no continuous optimization algorithm exists in the literature that obtains a fractional approximate point for this relaxation.

\revcolor{To attack the problem in a different way, we employ a three step strategy. First, we change our focus to the underlying combinatorial structure of the problem, namely, permutations, by relaxing them into the space of fractional solutions. We identify feasibility constraints for these fractional solutions, so that they can be interpreted as allocation probabilities of a randomized policy that returns a valid permutation. Then, we define a linear programming relaxation of the original problem over the polytope of fractional feasible solutions. This relaxation entails exponentially many variables \emph{and} constraints. Second, we provide a polynomial-size linear programming relaxation of the aforementioned linear program. Third, we show how to round the optimal fractional solution of the latter linear program using swap rounding. The rest of this section is dedicated to a more detailed discussion of these steps. The reader can directly refer to \Cref{revenue} for an in-depth discussion of our algorithm.}

\revcolor{\noindent\textbf{Step (i) - the polytope of implementable fractional solutions:}} We characterize the polytope of feasible policies for ranking the elements sequentially in \Cref{sec:polytope}. More specifically, define a feasible policy for the ranking problem to be a procedure that starts with an empty list and keeps adding elements one by one at the end of the list until it ends up with a permutation. The choice of the elements at every step can be deterministic or randomized. Given a randomized permutation generated by a feasible policy, one can define a collection of probability distributions, one for each $i\in[n]$, where the probability distribution corresponding to $i\in[n]$ is the induced distribution over sets of size $i$ by the first $i$ elements in that permutation. Notably, not all collections of distributions are \emph{implementable} by some randomized permutation. Therefore, we first identify necessary and sufficient conditions for the implementability of such a collection, which gives us the implementable polytope. Unfortunately, the description of this polytope  requires an exponential number of variables and (doubly) exponential number of constraints. 


\revcolor{\noindent\textbf{Step (ii) - relaxing the polytope and approximating the relaxation:} To overcome the above obstacle, in \Cref{sec:bicriteria} we first introduce a simpler relaxation of the aforementioned exponential-sized linear program that has polynomial number of constraints. In this relaxation, among  all the original constraints of the polytope, we only preserve a quadratic size subset. Roughly speaking, the constraints of the relaxed program correspond to the vector of probabilities of elements appearing at different positions being a feasible point in the Birkhoff-von Neumann perfect-matching polytope. We then approximately solve this relaxation linear program using the ellipsoid method~\citep{khachiyan1979polynomial,bubeck2015convex} by incorporating an approximate separation oracle for its dual LP --- which has polynomial number of variables and exponential number of constraints. }

\revcolor{\noindent\textbf{Step (iii) - rounding the fractional approximate solution:} For rounding, which is necessary to obtain a randomized permutation, we consider the so called \emph{marginals} of the above fractional approximate solution, which correspond to the probabilities of different elements appearing in different positions. We show that these marginals provide a feasible solution in the base polytope of the laminar matroid used in our reduction for Problem~\ref{pbm:seq}. Hence, we can input these marginals to the swap rounding randomized algorithm, so as to return a base of the laminar matroid. We finally use the post-processing step in the reduction to return a (randomized) permutation given a randomized base.}

Recall the submodular function in the larger space of size $n^2$ used in the reduction for Problem~\ref{pbm:seq}. To analyze the above algorithm, we consider one such function for each of the objective and group constraints. Also, we consider a randomized set $R^{\textrm{ind}}$ in this larger space of size $n^2$ where the $j$th copy of each element $i$ is \emph{independently} available in this set with probability equal to the corresponding marginal. Based on the guarantee of swap rounding~\citep{vondrak2011submodular,chekuri2014submodular} and that it is \emph{oblivious} to the choice of the underlying submodular function, the expected value of each of these submodular functions can only increase at the output randomized base, compared to the expected value at the randomized set $R^{\textrm{ind}}$. \revcolor{We then bound the final approximation factor using the correlation gap for submodular functions~\citep{agrawal2010correlation}, that compares the value of submodular functions on $R^{\textrm{ind}}$ versus any other randomized set with the same marginals.} This approach yields a bi-criteria $\left((1-1/e)^2, (1-1/e)^2\right)$-approximation algorithm for Problem~\ref{pbm:grp} up to an arbitrarily small additive error.

\revcolor{Due to the existence of a $(1-1/e)$-approximation algorithm for Problem~\ref{pbm:seq}, it is natural to ask whether a bi-criteria $(1-1/e,1-1/e)$-approximation algorithm exists for Problem~\ref{pbm:grp}. }
The main barriers in the above approach to obtain such an approximation ratio for Problem~\ref{pbm:grp} are (i) obtaining a tractable relaxation that can be solved with no or small approximation loss and (ii) designing an exact or approximate oblivious randomized rounding for that relaxation. In \Cref{appCoverage}, we show how to overcome these barriers for a practical special case of the problem. In particular, when the underlying submodular functions in the objective and constraints are \emph{coverage functions} -- for example, as in the most general scenario in ~\cite{ferreira} -- we obtain an alternative $(1-1/e)$-approximation algorithm for Problem~\ref{pbm:seq} and an improved bi-criteria $\left(1-1/e,1-1/e\right)$-approximation algorithm for Problem~\ref{pbm:grp}. Our algorithms rely on a different LP relaxation under coverage functions. Moreover, the new algorithms use a simple independent randomized rounding scheme that is oblivious to the choice of submodular functions and only loses at most $1/e$ fraction of the value of the sequential submodular functions appearing in the objective and group constraints.

 \subsection{Related Work}\label{literature_review}

From a methodological point of view, our work fits within the literature of submodular optimization \citep{calinescu2011maximizing,feige2011maximizing,chekuri2014submodular,buchbinder2015tight, sviridenko2017optimal,niazadeh2018optimal}. \revcolor{The main technical challenge of our work is due to selecting permutations instead of subsets. The same challenge is also present in the line of work on maximizing \emph{``sequence submodular functions''}~\citep{tschiatschek2017selecting,alaei2021maximizing,mitrovic2018submodularity}, which is another generalization of submodular maximization to the space of permutations. In this model, a directed acyclic graph (DAG) defined over the elements and a submodular function defined over the edges of this DAG are given. Then, the corresponding sequence submodular function assigns a value to each ranked subset of elements by first identifying the subset of edges of the DAG that are oriented consistently with this ranked subset, and then inputting this subset of edges to the submodular function. Although marginally related, to our knowledge there is no formal mathematical connection between this line of work and our problem, both in terms of models and algorithms.}
Also, our paper is related to the literature on submodular welfare maximization~\citep{vondrak2008optimal,mirrokni2008tight}. This problem is a special case of submodular maximization with a partition matroid constraint. Although our approximation algorithm for Problem~\ref{pbm:seq} is also based on a reduction to submodular maximization with a (laminar) matroid constraint, it needs to return a permutation at the end and ends up being very different from the reduction used for submodular welfare maximization.


From an application point of view, our work is related to assortment planning, which is the study of optimally presenting a subset of products to a user. Assortment optimization had an extensively growing literature over recent decades. We refer the reader to  related surveys and books \citep[cf.][]{KFV-08, lan-90, HT-98} for a comprehensive study. In particular, various consumer choice models have been considered in the literature, for example, multinomial logit models \citep{TV-04, LV-08, top-13}, Markov chain choice models~\citep{blanchet2016markov,feldman2017revenue}, ranked-list preference \citep{HGS-10, GLS-16}, and non-parametric (data-driven) choice models \citep{FJS-13,jagabathula2017nonparametric}. \revcolor{Our work diverges from the classic assortment optimization literature in that we optimize over the space of permutations rather than subsets.} 

\revcolor{There is also a rich literature on product ranking optimization in the revenue management literature~\citep{abeliuk2016assortment,gallego2020approximation,aouad2021display,sumida2021revenue}, which focus on maximizing revenue (i.e., product-weighted objectives) or a combination of revenue and social welfare, rather than the total purchase rate. Thus, the property of submodularity is not an “obvious” structure in these settings and our techniques cannot be directly applied to these problems. See \Cref{subsec:revenue} for an extension of our main result to the revenue management setting.}

A work closely related to ours is that of \cite{ferreira} which considers the problem of ranking assortments in the context of online retail, with the objective of maximizing user engagement. This problem is a special case of sequential submodular optimization, where the submodular functions of interest are all coverage functions. \cite{ferreira} provides the first constant approximation for the model they introduce by showing greedy algorithm is $0.5$-approximation, while also showing it is $(1-1/e)$-approximation under the assumption that click probabilities and patience levels are independent. They eventually feed greedy to a ``learning-then-earning''algorithm under i.i.d. stochastic offline learning setting to obtain vanishing approximate error bounds.

Our work is also  related to display advertising and sponsored search ad auction. The platform's decision on which ads to show and in what order, in different pages of a website 
or within a given page, 
is closely related to the information the platform has access to, regarding both the browsing and clicking behavior of the users. A stream of work, both in marketing  (e.g. see \citealt{anand_shachar_2011} and \citealt{hoban_bucklin_2015}) and in optimization (e.g. see, \citealt{mcafee_2009}, \citealt{balseiro_2014}, \citealt{aouad2021display}, and \citealt{sayedi_2018}) study various aspects of this problem. In particular, the trade-off between engagement, 
revenue, and ad diversity 
in online advertising has been a topic of investigation  (e.g. see, \citealt{lehaei_pennock_2007}, \citealt{radlinski_et_al_2008} and \cite{agrawal2018proportional}). In case of display advertising, \cite{zhu_2009} provide a machine learning algorithm for jointly maximizing revenue while providing high quality ads. More recently, \cite{ilvento2020multi} study the design of sponsored search ad auctions with multi-category fairness. 

The positive effects of diversification on consumers' purchasing behavior have been empirically observed (e.g., see \citealt{manchanda2006} and \citealt{kannan2014}). From a modeling perspective, similar optimization problems involving group constraints defined on assignment problems have been studied in the context of display advertising by \cite{asadpour2019}. Moreover, such diversification effects can be utilized to avoid marginalizing a subset of users. \cite{celis2019controlling, celis2019toward} discuss this consideration, which is called group fairness, as mentioned before, in details. 

Another related work to ours is \cite{niazadeh2020online}, which considers the online learning version of our problem. They show the simple greedy algorithm for sequential submodular maximization (see \Cref{greedyanalysis}) can be converted into an online learning algorithm. This work poses the interesting question of whether or not our $(1-1/e)$-approximation algorithm can be turned into an online learning algorithm as well. Also important is the result of \cite{goli}, who also study product ranking in online platforms. Their semantics are different from ours in crucial ways; for example, they consider a substantially different shopper behavior, which makes the two models incomparable.  Finally, the particular choice model we consider in our paper, in which a consumer first picks an attention window of a particular size and then chooses a product, is indeed an special case of the consider-then-choose choice model studied in \cite{aouad2020assortment}. This paper diverges from ours as they consider assortment optimization for maximizing revenue instead of product ranking for maximizing probability of click, and under a more general setting. 
 \section{Sequential Submodular Maximization for   User Engagement}
\label{engagement}

In this section, we study the product ranking for maximizing user engagement. As we discussed in \Cref{sec:intro}, this model can be cast as an instance of Problem~\ref{pbm:seq}, so we aim to find an optimal approximation algorithm for this more general problem. As a recap, the well-studied monotone submodular maximization subject to a cardinality constraint $k$ is in fact a special case of Problem~\ref{pbm:seq}, because it can be reduced to this problem by setting $\lambda_k=1$ and $\lambda_{i}=0$ for $i \neq k$. Hence, no approximation ratio better than $1-1/e$ is achievable unless $\textrm{P=NP}$.



Let $F(\pi)$ denote the sequential submodular function in the objective of Problem~\ref{pbm:seq} for a given permutation $\pi$. 
A natural candidate algorithm for this problem is probably a naive greedy algorithm that iteratively picks the element with the maximum marginal gain to $F(.)$ as the next element in the ordering $\pi$. See \Cref{alg:greedy} in \Cref{greedyanalysis}. For some special cases of the sequential submodular maximization -- for example, for a special case of the product ranking problem studied in \citet{ferreira} -- the approximation ratio of the greedy algorithm is known to be $1-1/e$. See \Cref{rem:prod-chris}. Unlike these special cases, the approximation ratio of this algorithm for the general sequential submodular maximization problem (and also the general product ranking with coverage choice functions, which is also studied in \cite{ferreira}) turns out to be exactly $1/2$. The proof of the approximation ratio and the tightness example are presented in \Cref{greedyanalysis}. 


The main result of this section is an \emph{optimal} $(1-1/e)$-approximation algorithm for Problem~\ref{pbm:seq} (\Cref{algEngagement}). We achieve this result through a reduction in \Cref{thm2}, which allows us to study an equivalent submodular optimization over matroids. \revcolor{We emphasize that Algorithm~\ref{algEngagement} is randomized and our approximation result in \Cref{thm2} holds \emph{in expectation}.}
\vspace{-3mm}
\begin{algorithm}[ht]
\caption{An Optimal Approximation Algorithm for Maximizing User Engagement}
\begin{algorithmic}[1]
\State \textbf{Input}: $n, \{f_i(\cdot)\}_{i\in [n]}, \{\lambda_i\}_{i\in [n]}$; \textbf{Output}: a permutation $\hat{\pi}$.
\State Consider a ground set of elements $\mathcal{V}\triangleq\{i^j \text{ for all } i, j \in [n]\}$ and a laminar collection $\mathcal{A}\triangleq\{A_k\}_{k\in[n]}$ of subsets of $\mathcal{V}$, where $\forall k\in [n]:~A_k\triangleq\{i^j \mid i\leq k, 1\leq j\leq n\}$.
\State Define the laminar matroid $\mathcal{M}\triangleq (\mathcal{V}, \mathcal{I})$, where a set $\Pi\subseteq \mathcal{V}$ is in $\mathcal{I}$ if and only if $\lvert \Pi\cap A_k \rvert\leq k$ for all $k\in [n]$.
 \State Define the monotone submodular function $G:2^{\mathcal{V}}\rightarrow \mathbb{R}$ as
 \begin{align*}
 	G(\Pi)\triangleq\lambda_1 f_1(S^{\Pi,1})+\lambda_2 f_2(S^{\Pi,2})+\dots+\lambda_n f_n(S^{\Pi,n}),
 \end{align*}
where for every $\Pi\subseteq \mathcal{V}$ and position $i\in[n]$, $S^{\Pi,i}$ is defined as: $$S^{\Pi,i}\triangleq\{j\in[n] \mid \exists ~ k\leq i ~~\text{such that}~~  k^j \in \Pi\}.$$
\State Find a (randomized) set $\hat{\Pi}\in\mathcal{I}$ by applying Calinescu et al.'s  $(1-1/e)$-approximation algorithm for maximizing monotone submodular functions over matroids~\citep{calinescu2011maximizing} to  $\underset{\Pi\in \mathcal{I}}{\max}~{G(\Pi)}$.
\State \textbf{Post-processing}: For each $j\in [n]$, let $\ell(j,\hat{\Pi})$ be the smallest $i$ for which $i^j\in \hat{\Pi}$, and $n+1$ if no such $i$ exists. Sort the elements in increasing order of their $\ell(\cdot,\hat{\Pi})$ values (breaking ties arbitrarily) and call this permutation $\hat{\pi}$. 
\State Return $\hat{\pi}$.
\end{algorithmic}
\label{algEngagement}
\end{algorithm}




\begin{theorem}\label{thm2}
	 \Cref{algEngagement} is an $(1-1/e)$-approximation algorithm for the sequential submodular maximization problem as defined in Problem~\ref{pbm:seq}. No polynomial-time algorithm can obtain a better approximation ratio unless $\textrm{P=NP}$. 
\end{theorem}

\subsection{Analyzing \Cref{algEngagement}}
\label{sec:analyzing-main-alg}
We first recap some preliminary notations and results that are needed to prove \Cref{thm2}. We reduce the problem to submodular maximization subject to a laminar matroid, for which we know -- due to \cite{calinescu2011maximizing} -- an optimal $(1-1/e)$-approximation exists.

\paragraph{(I) The laminar matroid.} The  underlying laminar matroid $\mathcal{M}(\mathcal{V}, \mathcal{I})$ is as follows. Ground set $\mathcal{V}$ contains $n^2$ elements $i^j$ for all $i ,j \in [n]$, where each element $i^j$ corresponds to placing element $j$ of the original problem in position $i$ in the permutation. The collection of independent sets $\mathcal{I}$ is defined by a laminar family  $\mathcal{A}$ and a capacity function $c:\mathcal{A}\rightarrow\mathbb{R_+}$, such that a set $\Pi\subseteq \mathcal{V}$ is in $\mathcal{I}$ (that is, $\Pi$ is an independent set of the laminar matroid) if and only if $|\Pi \cap A| \leq c(A)$ for each $A \in \mathcal{A}$. 
Here, the laminar family $\mathcal{A}$ is $\{A_1, A_2, \dots, A_n\}$, where
$A_k=\{i^j \mid i\leq k, 1\leq j\leq n\},$ and $c(A_k)$ is equal to $k$.

\revcolor{
\begin{remark}
Instead of defining a laminar matroid, one may be tempted to directly optimize $g(\cdot)$  over the space of subsets that correspond to permutations. More specifically, we can treat each element $i^j\in \mathcal{V}$ as an edge between position $i$ and item $j$ and then maximize $g(\cdot)$ over the space of perfect matchings. Importantly, this approach fails to obtain the optimal bound due to the impossibility of a particular lossless rounding in the perfect matching polytope. See \Cref{sec:appAlternate} for details.
\end{remark}}

\paragraph{(II) The monotone submodular function in the larger space.} Given $f_1(\cdot), f_2(\cdot), \dots f_n(\cdot)$, the function $G:2^\mathcal{V}\rightarrow\mathbb{R}_+$ is defined for each $\Pi \subseteq \mathcal{V}$ as
$$G(\Pi)\triangleq\lambda_1 f_1(S^{\Pi,1})+\lambda_2 f_2(S^{\Pi,2})+\dots+\lambda_n f_n(S^{\Pi,n}),$$
where $S^{\Pi,i}\triangleq\{j\in[n] \mid \exists ~ k\leq i ~~\text{such that}~~  k^j \in \Pi\}$.

First, we establish the main properties of function $G$. 
\begin{proposition}\label{proposition1}
 If $\{f_i\}_{i\in[n]}$ are monotone submodular functions in the ground set $[n]$, then the function $G(\Pi)=\sum_{i\in[n]}\lambda_i f_i(S^{\Pi,i})$ is monotone and submodular in the ground set $\mathcal{V}$. 
\end{proposition}
\begin{proof}{Proof.}
	To prove this proposition, we need to show that for any two subsets $\Pi, \Pi' \subseteq \mathcal{V}$ such that $\Pi\subseteq \Pi'$, the following two properties hold:
	\begin{enumerate}[label=(\roman*)]
		\item $G(\Pi)\leq G(\Pi')$.
		\item $G(\Pi\cup \{i^j\})-G(\Pi)\geq G(\Pi'\cup \{i^j\})-G(\Pi')$ for any $i^j\in \mathcal{V}$.
	\end{enumerate}
	To do this, let 
	\begin{align*}
		G(\Pi)=\sum_{i\in[n]}\lambda_i f_i(S^{\Pi,i})~~,~~
		G(\Pi')=\sum_{i\in[n]}\lambda_i f_i(S^{\Pi',i}).
	\end{align*}
	By definition, $S^{\Pi,i}$ is indeed the set of all elements in $[n]$ that appear at least once at a position no later than $i$ in $\Pi$. So we must have $S^{\Pi,i}\subseteq S^{\Pi',i}$ for all $i\in [n]$. Because each $f_i$ is monotone, we have $f_i(S^{\Pi,i})\leq f_i(S^{\Pi',i})$ for $i\in [n]$.
	This proves property (i). To see property (ii), note that for any $i^j\in \mathcal{V}$,
	\begin{align*}
		G(\Pi\cup \{i^j\})-G(\Pi)=\lambda_i \big(f_i(S^{\Pi,i}\cup \{j\})-f_i(S^{\Pi,i})\big)+\dots+\lambda_n \big(f_n(S^{\Pi,n}\cup \{j\})-f_n(S^{\Pi,n})\big),\\
		G(\Pi'\cup \{i^j\})-G(\Pi')=\lambda_i \big(f_i(S^{\Pi',i}\cup \{j\})-f_i(S^{\Pi',i})\big)+\dots+\lambda_n \big(f_n(S^{\Pi',n}\cup \{j\})-f_n(S^{\Pi',n})\big),
	\end{align*}
	Again, because each $f_{\ell}$ is submodular, we have for each $\ell\in [i:n]$
	$$f_{\ell}(S^{\Pi,\ell}\cup \{j\})-f_\ell(S^{\Pi,\ell})\geq f_\ell(S^{\Pi',\ell}\cup \{j\})-f_\ell(S^{\Pi',\ell}),$$
	which proves the second property.\hfill\Halmos
\end{proof}

\paragraph{(III) The reduction.} Recall that $F(\pi)$ denotes the sequential submodular function in the objective of Problem~\ref{pbm:seq} for a given permutation $\pi$. We now show that the problem of maximizing $F(\pi)$ over the space of all permutations reduces (in polynomial time) to optimizing $G(\Pi)$ over the space $\mathcal{V}$ subject to $\Pi$ be one of the independent sets of the laminar matroid $\mathcal{M}$ defined earlier. We show this by first converting any feasible solution $\Pi\in \mathcal{I}$, potentially resulting from maximizing or approximately maximizing $G(\cdot)$ over independent sets of $\mathcal{M}$, into a set $\tilde{\Pi}\subseteq \mathcal{V}$ corresponding to a permutation $\pi$ such that $G(\tilde{\Pi}) \geq G(\Pi)$.  Then, we show the linear combination of $f_i(\cdot)$'s over  $\pi$  has the same value as $G(\tilde{\Pi})$, i.e., $F(\pi)=G(\tilde{\Pi})$, which proves the reduction. The final approximation guarantee is proved by further showing the optimal objective value of the former problem is no smaller than the latter problem, because any permutation $\pi$ can be naturally mapped to a set $\Pi\in\mathcal{I}$ such that $G(\Pi)=F(\pi)$. 

\revcolor{
\begin{lemma}\label{claim3}
	Given a set $\Pi\in \mathcal{I}$, we can create a permutation $\pi:[n]\rightarrow[n]$ such that $F(\pi)\geq G(\Pi)$.
\end{lemma}}
\begin{proof}{Proof.}
For each element $j\in [n]$, we define $\ell(j,\Pi)$ to be the smallest $i\in [n]$ such that $i^j\in \Pi$, and $n+1$ if no such $i$ exists. We then sort the elements in the increasing order of their $\ell(\cdot,\Pi)$ values (we arbitrarily break the ties) to get a permutation $\pi$. We claim $F(\pi) \geq G(\Pi)$. To see this, consider an element $j$. Because $\Pi\in \mathcal{I}$, we must have
$$\sum_{k=1}^n\sum_{i=1}^{\ell(j,\Pi)}  \mathbb{I}\{{i^k\in \Pi}\} \leq \ell(j,\Pi)$$
where $\mathbb{I}\{{i^k\in \Pi}\}$ is the indicator variable that has a value of $1$ if $i^k\in \Pi$, and $0$ otherwise. This inequality implies that the position at which each element $j$ appears in $\pi$, denoted by $\pi^{-1}_j$, has to be less than or equal to $\ell(j,\Pi)$. Also, by definition, we have $\{\pi_1,\ldots,\pi_i\}=\{j\in[n]:\pi^{-1}_j\leq i\}$ and $S^{\Pi,i}=\{j\in[n]:\ell(j,\Pi)\leq i\}$ for each $i\in[n]$. So we have $\{\pi_1,\ldots,\pi_i\}\supseteq S^{\Pi,i}$, and hence $ f_i\left(\{\pi_1,\ldots,\pi_i\}\right)\geq f_i(S^{\Pi,i})$, for every $i\in[n]$. Therefore, by definition, we must have $F(\pi) \geq G(\Pi)$. \hfill\Halmos
\end{proof}
\revcolor{
\begin{lemma}\label{claim2}
Let $\mathcal{S}_n$ be the $n$-permutation group. Then $\underset{\Pi\in \mathcal{I}}{\max}~G(\Pi) \geq \underset{{\pi \in \mathcal{S}_n}}{\max}~F(\pi)$.
\end{lemma}}
\begin{proof}{Proof.}
	Suppose $\bar{\pi}=\underset{\pi\in \mathcal{S}_n}{\argmax}~F(\pi)$.  Let $\bar{\Pi}=\{1^{{\bar{\pi}}^{-1}_1}, 2^{{\bar{\pi}}^{-1}_2}, \dots, n^{\bar{\pi}^{-1}_n}\}$, where $\bar{\pi}^{-1}_j$ is the position of element $j$ in permutation $\bar{\pi}$ (or equivalently, $\bar{\pi}_i=j$ if and only if $i=\bar{\pi}^{-1}_j$). 
	Note that $\bar{\Pi}\in \mathcal{I}$, as there are exactly $i$ elements in positions $k\leq i$ in $\bar{\Pi}$. Moreover, by definition, $G(\bar{\Pi})=F(\bar{\pi})$, which finishes the proof. \hfill\Halmos
\end{proof}

Note that the optimal permutation maximizing $F(\pi)$ is indeed deterministic. Now we are ready to prove \Cref{thm2} by putting together the preceding results. 

\begin{proof}{Proof of \Cref{thm2}.}
Due to \Cref{proposition1} and  \cite{calinescu2011maximizing}'s $(1-1/e)$-approximation algorithm for maximizing a monotone submodular function subject to matroid constraints, we can find a randomized set $\hat{\Pi}\in \mathcal{I}$ in polynomial time such that $\expect{G(\hat{\Pi})}\geq (1-1/e) \max_{\Pi\in \mathcal{I}} G(\Pi)$. Combining this with \Cref{claim2} shows $\expect{G(\hat{\Pi})}\geq (1-1/e) \max_{\pi\in \mathcal{S}_n} F(\pi)$. Finally, by \Cref{claim3}, we can turn every realization of the set $\hat{\Pi}$ into a permutation $\hat{\pi}$ (the same one created by \Cref{algEngagement}) such that 
$$\expect{F(\hat{\pi})} \geq \expect{G(\hat{\Pi})}\geq (1-1/e) \max_{\pi\in \mathcal{S}_n} F(\pi),$$ which finishes the proof of the first part of the theorem.

To prove the hardness, as mentioned earlier, note that for the special case where all the functions $f_i$ are coverage functions, this problem reduces (in an approximation-preserving way) to a maximum coverage problem for which a $(1-1/e)$-hardness of approximation is known. Therefore, unless $\textrm{P=NP}$, no polynomial time algorithm can achieve a better than $(1-1/e)$-approximation. \hfill\Halmos
\end{proof}


\begin{remark}
\label{rem:prod-chris}
As stated in the proof of \Cref{thm2}, no policy can achieve a better than $(1-1/e)$-approximation even when $f_i(\cdot)$'s are coverage functions. Coverage functions have appeared in important special cases of our problem studied in \cite{ferreira}. They consider the product ranking problem with a particular choice function and independent patience levels, which results in a sequential submodular maximization problem for a special coverage function. For general coverage functions, we obtain an alternative $(1-1/e)$-approximation using an LP-based approach, which might be of independent interest. We present this result and its extension to Problem~\ref{pbm:grp} in  \Cref{appCoverage}.
\end{remark}

\revcolor{
We end this section by two remarks. First, while the focus of our paper is on maximizing user engagement (or equivalently purchase rate), the literature on assortment planning and revenue management mostly focuses on maximizing expected revenue. Unfortunately, the submodular structure of the objective function will not hold anymore if we consider heterogeneous revenues -- even for the special case of assortment planning with cardinality constraints and under random utility models such as MNL~\citep{desir2015capacity}. However, in \Cref{subsec:revenue} we show how to use a simple thresholding algorithm in order to get an approximation ratio for the revenue optimization problem by using \Cref{algEngagement} as a black box. Second, we briefly discuss going beyond submodularity assumption in \Cref{subsec:beyondSM} and comment on the flexibility of our approach in Problem~\ref{pbm:seq} --- which is almost using functions $f_1,\ldots,f_n$ in blackbox fashion as long as they satisfy certain properties.}




\section{ Sequential Submodular Maximization with Group Constraints}\label{revenue}
Recall the product ranking problem with group fairness constraints: given $L$ user groups, the goal is to find a (randomized) permutation $\pi$ over $n$ elements to  (i) maximize the overall user engagement, and (ii) guarantee the engagement of each group $l$ is not below a given threshold $T_l$. Indeed, this problem is an instance of Problem~\ref{pbm:grp}, where we have $L$ different sequential submodular group constraints, one for each group, with lower-bounds $\{T_l\}_{l\in[L]}$.

\revcolor{To understand a fundamental limitation of algorithms in Problem~\ref{pbm:grp} and the usefulness of randomness, consider its (equivalent) feasibility version where there are $L$ sequential submodular group constraints that need to be satisfied (in expectation, if the solution is randomized).\footnote{The optimization and the feasibility versions of Problem~\ref{pbm:grp} are equivalent, as by knowing the optimal value we can remove the objective function and add an extra constraint asking the value of the objective function to be at least the optimal value. The reduction is complete by searching over the optimal objective value using binary search.}
Now consider the following simple example: suppose we have two products $i_1$ and $i_2$ and two types of users $u_1$ and $u_2$, both with a deterministic patience equal to one (meaning that they only look at the first item). The user type $u_1$ only clicks on product $i_1$ and $u_2$ only click on product $i_2$, i.e., $$\kappa_{u_1}(S)=\mathbb{I}\{i_1\in S\},~\kappa_{u_2}(S)=\mathbb{I}\{i_2\in S\}$$
Suppose we have a separate group constraint for each type with thresholds $T_1=T_2=0.5$. Clearly, these  are sequential submodular group constraints, as the above functions are submodular. Note that there is no permutation over products $\{i_1,i_2\}$ where both types receive non zero engagement. However if we output $(i_1, i_2)$ or $(i_2, i_1)$ uniformly at random, then both types will receive an engagement of $0.5$ in expectation. As this example suggests, no deterministic solution can be approximately feasible in this problem. Recall that in our definition of Problem~\ref{pbm:grp}, we allow randomized solutions and they only need to be feasible in expectation. In such a setting, to have a framework when designing approximation algorithms, we focus on solutions that are approximately feasible in expectation, or equivalently we consider in expectation \emph{bi-criteria (approximation) guarantees} for the optimization version.}


\begin{definition}[Bi-criteria Approximation Ratio]
Suppose Problem~\ref{pbm:grp} has a feasible solution and let $D_\pi^*$ denote the optimum permutation distribution for this problem. An algorithm is a bi-criteria $(\alpha, \beta)$-approximation if it finds a permutation distribution $\hat{D}_{\pi}$ where 
\begin{align*}
\begin{array}{lll}
   \mathbb{E}_{\pi\sim \hat{D}_\pi}\left[ \displaystyle\sum_{i = 1}^n \lambda_i f_i(\{\pi_1, \pi_2, \ldots, \pi_i\})\right]&\geq \alpha\cdot \mathbb{E}_{\pi\sim {D}^*_\pi}\left[\displaystyle\sum_{i = 1}^n \lambda_i f_i(\{\pi_1, \pi_2, \ldots, \pi_i\})\right]&\\[20pt]
     \mathbb{E}_{\pi\sim \hat{D}_\pi}\left[\displaystyle\sum_{i = 1}^n \lambda^l_i f^l_i(\{\pi_1, \pi_2, \ldots, \pi_i\})\right]&\geq \beta \cdot T_l&\forall l\in[L]
\end{array}
\end{align*}
\end{definition}

The rest of this section is organized as follows. In \Cref{sec:polytope}, we characterize the set of feasible (randomized) policies for ranking elements as a polytope with exponentially many variables and constraints. By using a relaxation of this polytope in \Cref{sec:bicriteria},  we give a
bi-criteria $\left((1-1/e)^2, (1-1/e)^2\right)$-approximation algorithm for Problem~\ref{pbm:grp}.



\subsection{Polytope of Feasible Ranking Policies}
\label{sec:polytope}



Define a \emph{feasible policy} for ranking of the elements to be a procedure that starts with an empty list and keeps adding elements one by one to the end of the list until it ends up with a permutation. The choice of the elements at every step can be deterministic or randomized, and it can be adaptive or oblivious. Note that any such policy can be identified as a distribution over the space of permutations. 

For every $i\in [n]$ and $S\subseteq [n]$ with $|S|=i$, let $x_{i, S}$ represent the probability that the set of the first $i$ elements in the permutation is $S$. We say an assignment of values to $\vec{x}$ is \emph{implementable} if a feasible policy exists such that for every set $S\subseteq [n]$ of size $i$, the probability that it places the elements in set $S$ in the first $i$ positions is exactly $x_{i, S}$. 
We next identify the necessary and sufficient conditions for implementability of $\vec{x}$ in \Cref{condition}. See \Cref{sec:proof-of-polytope} for the proof of this proposition.

\begin{proposition}\label{condition}
	The vector $\vec{x}$ is implementable by a feasible policy, if and only if
	\begin{enumerate}[(i)]
	    \item \label{cond:i} For every $i\in [n]$ and $S\subseteq [n]$ with $|S|=i$, we have $x_{i, S} \geq 0$.
		\item \label{cond:ii} For each $1\leq i\leq n$, we have $\sum_{S\subseteq [n], |S|=i} x_{i, S}=1$.
		\item \label{cond:iii} For any collection $\mathcal{C}$ of subsets of $[n]$ with size $1\leq i < n$, we have:
$$\sum_{S\in \mathcal{C}} x_{i, S}\leq \sum_{T\in N(\mathcal{C})} x_{i+1, T},$$
where $N(\mathcal{C})$ is the collection of subsets  $T\subseteq[n]$ of size $i+1$ such that there exists set $S\in \mathcal{C}$  (with size $i$) so that $S\subset T$.
	\end{enumerate}
\end{proposition}
Given \Cref{condition}, we can define the polytope of feasible policies, denoted by \ref{feasible-poly}, as follows:
\begin{align*}\label{feasible-poly}
\tag{$\mathcal{P}$}
 &\sum_{S\subseteq [n], |S|=i} x_{i, S}=  1, &  \forall i\in[n]\\
&\sum_{S\in N(\mathcal{C})} x_{i, S}-\sum_{S\in \mathcal{C}} x_{i-1,S}\geq 0, &  \forall i\in[n], \mathcal{C}\in \mathscr{C}_i \\
 &x_{i, S} \geq 0,~~~~x_{0, \{\emptyset\}}=0, & \forall i\in [n], S\subseteq [n], |S|=i \nonumber 
\end{align*}
where $\mathscr{C}_i$ is the power set of all the subsets of $[n]$ with size $i$. Note that this polytope is characterized by exponentially many variables and (doubly) exponentially many constraints. 
\subsubsection{More on Implementability and the Proof of \Cref{condition}}
\label{sec:proof-of-polytope}

We first show the easy direction of \Cref{condition}, that is, if the values of the assignment $\{x_{i, S}\}_{S\subseteq[n]}$ correspond to probabilities derived from a feasible ranking policy, then it satisfies the conditions in \Cref{condition}. Condition~\ref{cond:i} is satisfied trivially, as probabilities are non-negative. Moreover, every policy by definition fills all the positions with some element, so a set $S$ of size exactly $i$ exists that is placed in the first $i$ positions. Thus, it satisfies condition~\ref{cond:ii}. For condition~\ref{cond:iii}, consider a collection $\mathcal{C}$ of subsets of $[n]$ with size $i$. Suppose $\pi$ is the randomized permutation of the policy. Note $N(\mathcal{C})=\{S\cup\{j\}:S\in\mathcal{C},j\notin S\}$. Therefore, 
$$
\sum_{S\in \mathcal{C}}x_{i,S}=\mathbb{P}\left[\{\pi_1,\ldots,\pi_i\}\in\mathcal{C}\right]\leq \mathbb{P}\left[\{\pi_1,\ldots,\pi_i,\pi_{i+1}\}\in N(\mathcal{C})\right]=\sum_{T\in N(\mathcal{C})}x_{i+1,T}~.
$$


For the opposite direction, suppose $\vec{x}$ satisfies conditions \ref{cond:i}, \ref{cond:ii}, and \ref{cond:iii}. We show how to construct a feasible policy with assignment probabilities equal to $\vec{x}$. To do so, observe that we can check implementability of $\vec{x}$ inductively, i.e.,  ``layer by layer.'' Suppose there is a feasible policy for ranking the first $i$ elements, such that for every $S\subseteq [n]$ with cardinality $j\leq i$, $x_{j, S}$ be the probability that the set of first $j$ elements is $S$. We want to check if this policy can be extended in a consistent way to $i+1$, i.e., whether adding an additional element at position $i+1$ is possible in a way that for every $S\subseteq [n]$ with cardinality $i+1$, $x_{i+1, S}$ be the probability that the set of the first $i+1$ elements is $S$. We call this property \emph{implementability of layer $i+1$}.  From the definition, we clearly see that if $\vec{x}$ is implementable, all the layers $1\leq i \leq n$ are implementable as well. In addition, if all the layers $1\leq i \leq n$ are implementable, then $\vec{x}$ is implementable. 

\begin{figure}[htb]
\centering
\includegraphics[width=0.8\textwidth]{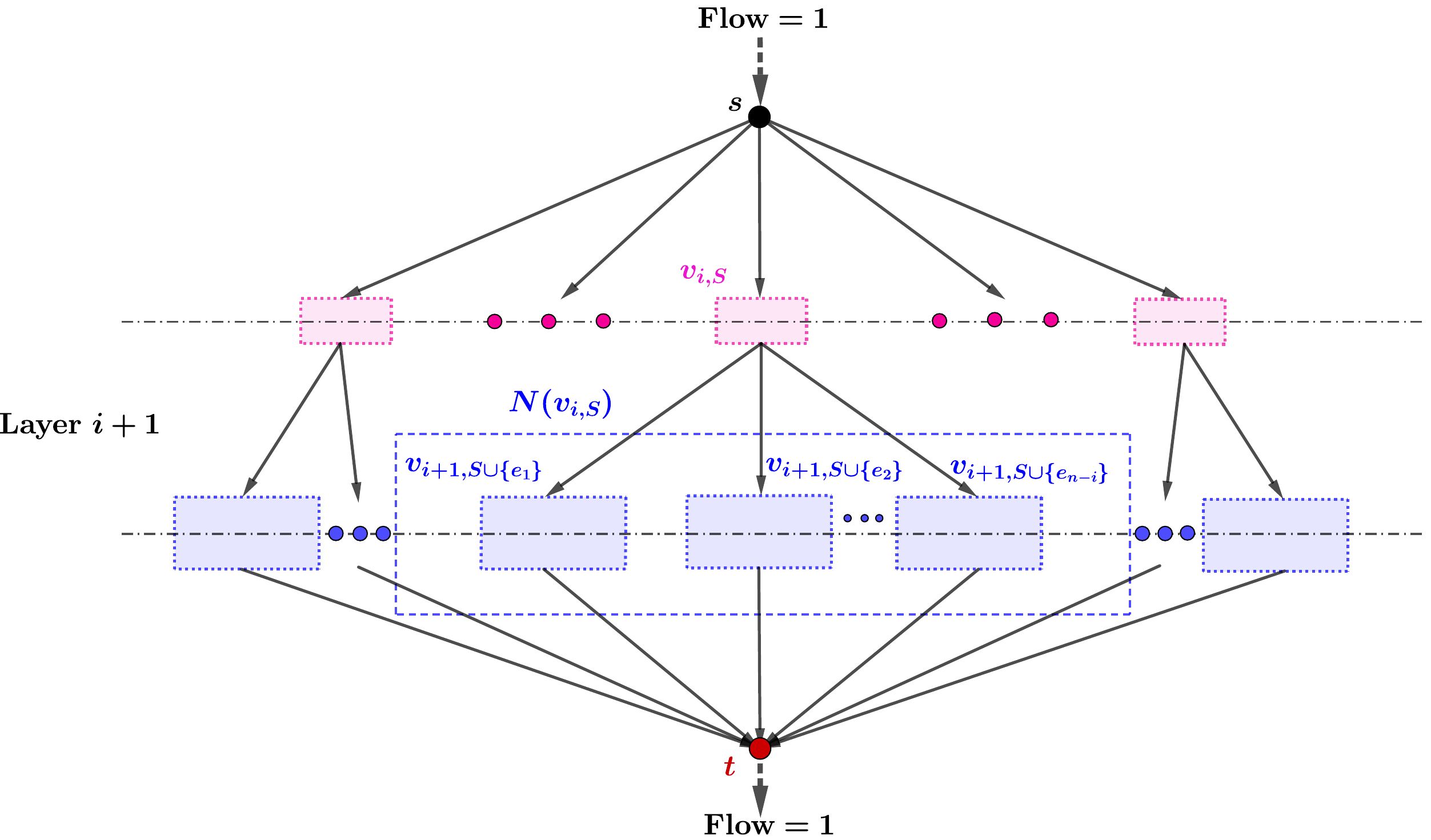}
\caption{\label{fig:flow} Network flow $G$ for layer $i+1$ (proof of \Cref{condition}). Capacity of an edge from source $s$ to node $S$ with $\lvert S\rvert=i$ is $x_{i,S}$,  from node $T$ with  $\lvert T\rvert=i+1$ to sink $t$ is $x_{i+1,T}$, and all inner-layer capacities are equal to $1$.}
\end{figure}

In order to check the implementability of a layer, we use a max-flow argument. Construct a flow network $G$, consisting of a node $v_{i, S}$ for each subset $S\subseteq [n], |S|=i$ and a node $v_{i+1, S}$ for each subset $S\subseteq [n], |S|=i+1$. For any two subsets $S\subset T$ of size $i$ and $i+1$, respectively, an edge from $v_{i, S}$ to $v_{i+1, T}$ with capacity $1$ exists. We also have a source $s$ and a sink $t$. The source is connected to each node $v_{i, S}$ with capacity $x_{i, S}$ and each node $v_{i+1, T}$ is connected to the sink with capacity $x_{i+1, T}$. Note the sum of the capacity of edges exiting the source and entering the sink are both $1$ due to condition (i). See \Cref{fig:flow} for more details. We now have the following equivalent characterization of implementability of a layer. Proof is postponed to \Cref{apx:proof-of-implement}.  

\begin{lemma}
\label{lemma:implement}
The layer $i+1$ is implementable if and only if a flow of $1$ from source node $s$ to sink node $t$ exists in the flow network $G$ (\Cref{fig:flow}). 	
\end{lemma}

Finally, given conditions~\ref{cond:i} and \ref{cond:ii}, we show a flow of $1$ from $s$ to $t$ at each layer exists if and only if condition~\ref{cond:iii} holds. Therefore, because of \Cref{lemma:implement}, all the layers are implementable. This last fact holds because of the generalized Hall's matching theorem for network flows~\citep{hall1935representatives}, or equivalent versions of the max-flow min-cut theorem~\citep{ford1958network}. \revcolor{ To see a simple standalone proof, in order to be able to send one unit of flow from $s$ to $t$ in $G$, all the cut sets separating $s$ from $t$ should have a value of at least $1$. Any cut set that crosses an edge from layer $i$ to $i+1$ has value at least $1$. Moreover, cut sets separating $s$ from the rest of the graph, or separating $t$ from the rest of the graph have values equal to $1$ because of condition~\ref{cond:ii}. Now consider a cut set separating $\{s\}\cup\mathcal{C}\cup N(\mathcal{C})$ from the rest of the nodes in $G$, where $\mathcal{C}$ is a collection of subsets of size $i$. The value of such a cut set, i.e., total capacity of crossed forward edges (minus backward edges) is equal to
$$
\sum_{S:\lvert S\rvert=i, S\notin \mathcal{C} }x_{i,S}+\sum_{T\in N(\mathcal{C})}x_{i,T}=1-\sum_{S\in \mathcal{C} }x_{i,S}+\sum_{T\in N(\mathcal{C})}x_{i,T}~,
$$
where the last equality holds because of condition~\ref{cond:ii}. Therefore, if condition~\ref{cond:iii} holds, the value of such a cut set is at least $1$, as desired.}

Given that all layers are implementable, we construct our final policy as follows. We start with an empty sequence, and at each step, we add a new element at the end of the current sequence.  If at step $i$ the set of added elements so far is $S$, at step $i+1$, we select element $\hat{p}\in [n]\setminus S$ independently from probability distribution $\left[\psi(S, S\cup \{p\})/x_{i, S}\right]_{p\in[n]\setminus S}$, and add it to the end of the sequence. Here, $\psi(S, S\cup \{p\})$ is the flow going from $v_{i, S}$ to $v_{i+1, S\cup \{p\}}$ in the graph we constructed to check implementability of layer $i+1$. Due to implementability of all layers from $1$ to $n-1$, this randomized policy implements the assignment $\vec{x}$, as desired.


\subsection{The Bi-criteria Approximately Optimal Policy}
\label{sec:bicriteria}

Recall the objective function of Problem~\ref{pbm:grp}. Given implementable $\vec{x}$, the objective value of the policy that implements $\vec{x}$ can be written as 
\begin{align}\label{revenueByx}
 \sum_{i \in [n]} \sum_{S\subseteq [n], |S|=i} \lambda_i x_{i, S}f_i(S).
\end{align}
Now, given the polytope of feasible policies \ref{feasible-poly}, the LP to find a randomized policy maximizing \eqref{revenueByx} while respecting the group constraints (i.e.,  keeping the generated user engagement above some threshold $T_l$ for each group $l \in [L]$) is
\begin{align*}
\max \quad
  &\sum_{i\in[n]}\sum_{S\subseteq [n], |S|=i} \lambda_ix_{i, S}f_i(S) & \text{s.t.}\nonumber \\
    &\sum_{i\in [n]}\sum_{S\subseteq [n], |S|=i} \lambda_i^l x_{i, S} f_i^l(S)\geq T_l, & \forall l\in [L]  \\
    &\left[x_{i, S}\right]_{i \in [n],S\in\mathcal{C}_i}\in\mathcal{P}. &  \nonumber
\end{align*}

 Unfortunately, the above linear program has exponentially many variables and (doubly) exponentially many constraints, which renders finding an exact solution through this LP computationally difficult. We are also not aware of any techniques to obtain an approximately optimal solution of this linear program directly. To circumvent these issues, we try to find a relaxation for this program that has exponentially many variables, but polynomially many constraints. We do so by dropping many of these constraints and only keeping polynomially many of them.  We then consider the dual of the relaxed program that has polynomially many variables and exponentially many constraints. Finally, we find an approximately optimal separation oracle for the dual, and by employing the ellipsoid method, we obtain an approximately optimal solution for the primal.



More formally, define $\mathcal{C}_{i, j}\triangleq\{S\subseteq [n] ~|~ |S|=i, j\in S\}$. Now, we relax the program by only considering constraints $\mathcal{C}\in \mathscr{C}_i$ that are among $\{C_{i, 1}, \ldots, C_{i, n}\}$. Define $y_{i,j}\triangleq\sum_{|S|=i, S\ni j} x_{i, S}-\sum_{|S|=i-1, S\ni j} x_{i-1, S}$. \revcolor{Note that intuitively, $y_{i, j}$ corresponds to the marginal probability of element $j$ appears in position $i$ according to $\vec{x}$.} Based on this definition of $y_{i,j}$, these constraints correspond to
$
y_{i,j}\geq 0 , \forall i,j\in[n]
$.
In doing so, we obtain the following relaxation of the previous linear program, denoted by \ref{LP1}:
\begin{align*}\label{LP1}\tag{\texttt{Primal}}
\max \quad
  & \sum_{i\in[n]}\sum_{S\subseteq [n], |S|=i} \lambda_i x_{i, S}f_i(S) &   \nonumber \\
 \text{s.t.} \quad  &y_{i, j} = \sum_{|S|=i, S\ni j} x_{i, S}-\sum_{|S|=i-1, S\ni j} x_{i-1, S}, &\forall i, j \in [n] \\
    &\sum_{i\in [n]}\sum_{S\subseteq [n], |S|=i} \lambda_i^l x_{i, S} f_i^l(S)\geq T_l, &  \forall l\in [L]\\
&\sum_{S\subseteq [n], |S|=i} x_{i, S}=  1, &  \forall i\in[n]\\
     &y_{i, j} \geq 0, & \forall i, j\in [n]\nonumber\\
    &x_{i, S} \geq 0,~~~~x_{0, \{\emptyset\}}=0.& \forall i\in [n], S\subseteq [n], |S|=i \nonumber
\end{align*}
\revcolor{
Note every implementable policy $\vec{x}$ is a feasible solution in the above LP relaxation: for such a policy, $y_{i,j}$ is equal to the probability that element $j$ is placed in position $i$. Therefore, the optimal objective value of \ref{LP1} is an upper-bound on the objective value of the optimal permutation distribution in Problem~\ref{pbm:grp}, i.e., the overall user engagement of the optimal implementable policy satisfying the group constraints.\footnote{In \Cref{sec:appendix-example}, we show by a computer-aided example that the objective value of \ref{LP1} might in fact be strictly larger than the user engagement of the optimal implementable policy.} We later use this connection when designing our rounding algorithm.}


In the rest of this section, we present a \revcolor{randomized algorithm for finding a permutation of elements that in expectation} achieves a constant fraction of $T_l$ as its user engagement for each group $l \in L$, and a constant fraction of the optimal objective value of \ref{LP1} as its overall user engagement. 
\revcolor{
\subsection{Detailed Steps of the Algorithm} 
\label{sec:alg-bicriteria-detail}
The first step in our proposed algorithm is to find an approximately optimal fractional solution to the relaxation linear program~\ref{LP1}. This linear program has exponentially many variables, but polynomially many constraints. To obtain an approximately optimal solution, we start by looking at its dual linear program, denoted by \ref{LP2}:
\begin{align} \label{LP2}\tag{\texttt{Dual}}
\min \quad
   &\sum_{i\in [n]} \alpha_i-\sum_{l\in [L]}\gamma_l T_l & \nonumber \\
    \text{s.t.} \quad  &\alpha_i + \sum_{j\in S} \beta_{i+1, j}-\sum_{j\in S} \beta_{i, j}\geq \lambda_i f_i(S)+\sum_{l\in [L]} \gamma_l \lambda_i^l f_i^l(S), &  \forall i\in [n-1], S\subseteq [n], |S|=i \label{firstConstraint} \\
    &\alpha_n -\sum_{j\in [n]} \beta_{n, j}\geq  \lambda_n f_n([n])+\sum_{l\in [L]} \gamma_l \lambda_n^l f_n^l([n]), & \nonumber\\
    &\beta_{i, j} \geq 0~~,~~\gamma_l \geq 0. & \forall i, j\in [n]~~,~~\forall l\in [L] \nonumber
\end{align}
Note that in the above linear program, dual variables $\{\beta_{i,j}\}$ correspond to the first set of constraints in \ref{LP1}, $\{\gamma_l\}$ correspond to the second set of constraints in \ref{LP1}, and $\{\alpha_i\}$ correspond to the third set of constraints in \ref{LP1}. Note also that this dual LP has exponentially many constraints (thanks to its first set of constraints), but polynomially many variables. While the  exact separation problem in this LP is likely to be computationally hard,\footnote{\revcolor{In fact, it is NP-hard for general monotone submodular functions $\{f^l_i\}$, as it generalizes the problem of maximizing the sum of negative linear functions and monotone submodular functions; see \cite{sviridenko2017optimal} for details.}} we show this LP is amenable to an approximate separation oracle. By carefully using this approximate separation oracle, we show the following proposition. Before stating proposition, we make a technical assumption regarding the boundedness of our submodular functions, to be able to handle small additive errors required in the separation oracle.~\footnote{\revcolor{This assumption is automatically satisfied when our submodular functions correspond to the purchase probability function of any given consumer choice model.}}
\begin{assumption}
\label{assum:bounded}
Submodular functions $\{f_i,f_i^{l}\}_{i\in [n],l\in[L]}$ are non-negative and bounded. Without loss of generality, we assume their range is $[0,1]$.
\end{assumption}

\begin{proposition}\label{mainLemma}
Fix $L$, $n$, $\{f_i(\cdot)\}_{i\in [n]}$, $\{\lambda_i\}_{i\in [n]}, \{f_i^l(.)\}_{i\in [n], l\in [L]}, \{\lambda_i^l\}_{i\in [n], l\in [L]}$, $\{T_l\}_{l\in [L]}$, and suppose Problem~\ref{pbm:grp} is feasible. For any $\epsilon>0$, there is an algorithm with polynomial running time in $\frac{1}{\epsilon}$, $n$ and $L$ (i.e., size of the problem instance) that generates an approximately optimal bi-criteria solution to \emph{\ref{LP1}} with additive error $O(\epsilon)$, that is, a fractional solution $(\hat{x},\hat{y})$ satisfying the following properties: 
	\begin{enumerate}
		\item This solution attains an objective value at least $(1-1/e)$ fraction of the optimal objective value of the linear program \emph{\ref{LP1}} minus $O(\epsilon)$,
		\item This solution satisfies the third constraint approximately; that is, for each group $l\in [L]$ we have
		$$\sum_{i\in [n]}\sum_{S\subseteq [n], |S|=i} \lambda_i^l \hat{x}_{i, S} f_i^l(S)\geq (1-1/e)T_l,$$
		and satisfies all the other constraints exactly.
	\end{enumerate}
\end{proposition}
}


\revcolor{The proof of \Cref{mainLemma} -- deferred to \Cref{sec:anlysis-main-bicriteria} --  is rather technical and based on constructing an algorithm; this proof relies on (i) using an approximation algorithm for a specific submodular maximization problem to play the role of the approximate separation oracle, (ii) then using this approximate separation oracle in the ellipsoid method (or any other cutting-plane method for solving LPs that can be endowed with an approximate separation oracle; see \cite{bubeck2015convex}, Chapter~2 for details) to solve a modified version of dual linear program \ref{LP2}, and finally (iii) showing how from this solution one can obtain an approximately optimal solution for \ref{LP1} in polynomial-time.}

Given access to the algorithm stated in \Cref{mainLemma}, we are now ready to describe our main bi-criteria approximation algorithm (\Cref{alg:bicriteria}).

\revcolor{
\begin{algorithm}[ht]
\caption{\label{alg:bicriteria}Approximation Algorithm for Maximizing Engagement Subject to Group Constraints}
\revcolor{
\begin{algorithmic}[1]

\smallskip
\State \textbf{Input}: $L, n, \{f_i(\cdot),\lambda_i\}_{i\in [n]}, \{f_i^l(.), \lambda_i^l\}_{i\in [n], l\in [L]}, \{T_l\}_{l\in [L]}$, $\epsilon$; \textbf{Output}: a permutation $\hat{\pi}$.



\State Apply \Cref{mainLemma} with $\epsilon>0$ to return an approximate bi-criteria solution $(\hat{x},\hat{y})$ to \ref{LP1}.

\State Consider a ground set of elements $\mathcal{V}\triangleq\{i^j \text{ for all } i, j \in [n]\}$ and a laminar collection $\mathcal{A}\triangleq\{A_k\}_{k\in[n]}$ of subsets of $\mathcal{V}$, where $\forall k\in [n]:~A_k\triangleq\{i^j \mid i\leq k, 1\leq j\leq n\}$.
\State Define the laminar matroid $\mathcal{M}\triangleq (\mathcal{V}, \mathcal{I})$, where  $\Pi\subseteq \mathcal{V}$ is in $\mathcal{I}$ if and only if $\lvert \Pi\cap A_k \rvert\leq k$ for all $k\in [n]$.

\State Run the swap randomized rounding algorithm~\citep{chekuri2010swap} for the laminar matroid $\mathcal{M}$, given $\hat{y}\in[0,1]^{n^2}$ as the input.  Let $\hat{\Pi}\in\mathcal{I}$ be the independent set that the algorithm returns.


\State \textbf{Post-processing}: For each $j\in [n]$, let $\ell(j,\hat{\Pi})$ be the smallest $i$ for which $i^j\in \hat{\Pi}$, and $n+1$ if no such $i$ exists. Sort the elements in increasing order of their $\ell(\cdot,\hat{\Pi})$ values (breaking ties arbitrarily) and call this permutation $\hat{\pi}$. 
\State Return $\hat\pi$.
\end{algorithmic}
}
\label{algRevenue}
\end{algorithm}
}


\begin{theorem}
\label{thm:main-bi}
	  Fix $L, n, \{f_i(\cdot)\}_{i\in [n]}, \{\lambda_i\}_{i\in [n]}, \{f_i^l(.)\}_{i\in [n], l\in [L]}, \{\lambda_i^l\}_{i\in [n], l\in [L]}$,  $\{T_l\}_{l\in [L]}$, and suppose Problem~\ref{pbm:grp} has a feasible solution. \revcolor{Then, given $\epsilon>0$,  \Cref{alg:bicriteria} finds a permutation $\hat\pi$ with polynomial ruuning time in $\frac{1}{\epsilon}$, $n$, and $L$  (i.e., size of the problem instance) such that
	  \vspace{-2mm}
	\begin{enumerate}
		\item The overall user engagement generated by $\pi$ is within factor $(1-1/e)^2$ of the objective value of \ref{LP1} minus $O\left(\epsilon\right)$; hence, it  is at least $(1-1/e)^2$ fraction (up to additive $O\left(\epsilon\right)$ error) of the overall user engagement of the optimal randomized policy satisfying the lower bounds $T_l$ on user engagement for each group.
		\item For each group $l\in [L]$, the user engagement generated by $\pi$ is at least $(1-1/e)^2T_l$.
	\end{enumerate}
	Therefore, it is a bi-criteria $\left((1-1/e)^2, (1-1/e)^2\right)$-approximation algorithm (up to an additive error in the order of $O(\epsilon)$), where $(1-1/e)^2\approx 0.4$. }
\end{theorem}


\begin{remark}
  We emphasize that \Cref{alg:bicriteria} is randomized and our bi-criteria approximation result in \Cref{thm:main-bi} holds \emph{in expectation}, both for the objective function and the group constraints.
\end{remark}

\revcolor{It is important to mention that \Cref{mainLemma} relies on an algorithm for approximately maximizing the sum of a monotone submodular function and a negative linear function from \cite{sviridenko2017optimal} as the separation oracle. As this algorithm is relying on continuous optimization methods, there is always an small (and negligible) additive error in its approximate performance guarantee.
\begin{remark}
  The algorithm in \cite{sviridenko2017optimal} suffers from a significant drawback, and that is the need for \emph{guessing} the contribution of the linear component of the objective to the optimal solution. As explained in the follow up work \cite{feldman2021guess}, this guessing can be problematic both as a computationally demanding pre-processing step, and also during the run of the modified version of continuous greedy algorithm in \cite{sviridenko2017optimal}. To circumvent these issues, \cite{feldman2021guess} proposes a clever modification to the continuous greedy algorithm that does not need the guessing step, and obtains exactly the same approximation guarantee as in \cite{sviridenko2017optimal}. To avoid the computational issues with guessing, we can instead use this algorithm in \Cref{mainLemma}.   
\end{remark}}




\vspace{-3mm}
\subsection{Analyzing \Cref{alg:bicriteria}}


\revcolor{As for the running time, all steps of \Cref{alg:bicriteria} are polynomial-time, and the algorithm used in \Cref{mainLemma} has a polynomial running time in $\frac{1}{\epsilon}$, $n$, and $L$ -- hence the desired running time. } Now let $(\hat{x},\hat{y})$ be the solution obtained by  \Cref{mainLemma} and let $\hat{T}_l$ be the user engagement generated by this solution for each group $l\in [L]$. Suppose there exists a policy that would show a permutation $\pi$ to users randomly, such that
$$\mathbb{P}[\{\pi_1, \ldots, \pi_i\}=S]=\hat{x}_{i, S} \hspace{0.5 in} \forall i\in [n], S\subseteq [n], |S|=i$$
Such a policy, if exists, would achieve $(1-1/e)$ fraction of the optimal objective of \ref{LP1} in expectation and would satisfy the lower-bound on the engagement of each group by a factor of $(1-1/e)$. \revcolor{However, the vector $\vec{\hat{x}}=\{\hat{x}_{i,S}\}$ is not necessarily a feasible point in the polytope $\mathcal{P}$ of feasible policies (as it is only a feasible point in a relaxation of $\mathcal{P}$); therefore, the  above policy might not exist.}

In the rest of our analysis we show how to devise a \emph{feasible} policy that achieves an approximation to $\vec{\hat{x}}$. As in the proof of \Cref{thm2}, we first define the laminar matroid $\mathcal{M}=(\mathcal{V}, \mathcal{I})$ such that
\begin{align*}
    &\mathcal{V}=\{i^j \text{ for all } i, j \in [n]\}~~~~~~,~~~~~\forall k\in [n]:~A_k=\{i^j \mid i\leq k, 1\leq j\leq n\}
\end{align*}
and a set $\Pi\subseteq \mathcal{V}$ is in $\mathcal{I}$ if and only if $\Pi\cap A_k \leq k$ for all $k\in [n]$. Similarly, given submodular functions $f_1, f_2, \dots, f_n$ in the objective of Problem~\ref{pbm:grp}, the function $G:2^{\mathcal{V}}\rightarrow \mathbb{R}$ is defined for each $\Pi \subseteq \mathcal{V}$ as
 \begin{align*}
 	G(\Pi)\triangleq\lambda_1 f_1(S^{\Pi,1})+\lambda_2 f_2(S^{\Pi,2})+\dots+\lambda_n f_n(S^{\Pi,n}),
 \end{align*}
where for every $\Pi\subseteq \mathcal{V}$ and position $i\in[n]$, $S^{\Pi,i}$ is defined as: $$S^{\Pi,i}\triangleq\{j\in[n] \mid \exists ~ k\leq i ~~\text{such that}~~  k^j \in \Pi\}.$$ Moreover, for each $l\in [L]$ and submodular functions $f^l_1, f^l_2, \dots, f^l_n$, the function $G^l:2^\mathcal{V}\rightarrow\mathbb{R}$ is defined for each $\Pi \subseteq \mathcal{V}$ as
	$$G^l(\Pi)\triangleq\lambda^l_1 f^l_1(S^{\Pi,1})+\lambda^l_2 f^l_2(S^{\Pi,2})+\dots+\lambda^l_n f^l_n(S^{\Pi,n}).$$
Note that functions $G$ and $\{G^l\}_{l\in[L]}$ are all non-negative monotone submodular functions from subsets of $\mathcal{V}$ to $\mathbb{R}$, thanks to \Cref{proposition1}. To move forward, we also need to use the following result due to \cite{agrawal2010correlation}, which bounds the \emph{correlation gap} of monotone submodular functions.
\begin{proposition}[\citealp{agrawal2010correlation}]\label{corrGap}
	Consider a monotone submodular function $f$ and a distribution $D$ over subsets $S\subseteq [n]$ such that $\mathbb{P}_{S\sim D}[i \in S]=p_i$. We have
	$$\frac{\mathbb{E}_{S\sim D^{\textrm{ind}}(\vec{p})}[f(S)]}{\mathbb{E}_{S\sim D}[f(S)]}\geq 1-1/e$$
where $D^\textrm{ind}(\vec{p})$ is the independent distribution with marginals $\vec{y}$, for which the probability of drawing a subset $S\subseteq [n]$ is
	$\displaystyle\underset{{i\in S}}{\Pi} p_i \underset{{i\notin S}}{\Pi} (1-p_i).$
\end{proposition}

Now suppose $\text{OPT}$ denotes the optimal objective value of \texttt{\textcolor{cornellred}{Primal}}. Inspired by \Cref{corrGap}, pick a randomized subset $\Pi^{\textrm{ind}}(\hat{y})\subseteq \mathcal{V}$ such that each element $i^j\in \mathcal{V}$ appears in $\Pi^{\textrm{ind}}(\hat{y})$ independently with probability $\hat{y}_{i, j}$. Because each variable $\hat{y}_{i, j}$ corresponds to the marginal probability of element $j$ appearing in position $i$ under the assignment  $\{\hat{x}_{i, S}\}$, due to \Cref{corrGap}, \revcolor{we have:
\begin{align}
\label{eq:third}
    \mathbb{E}\left[G\left(\Pi^{\textrm{ind}}(\hat{y})\right)\right]\geq (1-1/e) \sum_{i\in [n]}\sum_{S\subseteq [n], |S|=i} \lambda_i \hat{x}_{i, S} f_i(S) \geq (1-1/e)^2 \text{OPT}-O(\epsilon)~,
\end{align}
where the last inequality holds because of \Cref{mainLemma}. Also, for each $l\in[L]$ we have:
\begin{align}
\label{eq:fourth}
    \mathbb{E}\left[G^l\left(\Pi^{\textrm{ind}}(\hat{y})\right)\right]\geq (1-1/e) \sum_{i\in [n]}\sum_{S\subseteq [n], |S|=i} \lambda_i^l \hat{x}_{i, S} f_i^l(S)= (1-1/e) \hat{T}_l\geq (1-1/e)^2 T_l~,
\end{align}}
where again the last inequality holds because of \Cref{mainLemma}. Next, we claim for the randomized permutation $\pi$ returned by \Cref{alg:bicriteria}, the expected overall user engagement generated by $\pi$ is at least $\mathbb{E}\left[G\left(\Pi^{\textrm{ind}}(\hat{y})\right)\right]$ and the expected user engagement generated by $\pi$ for each group $l\in [L]$ is at least $ \mathbb{E}\left[G^l\left(\Pi^{\textrm{ind}}(\hat{y})\right)\right]$. This claim, together with \eqref{eq:third} and \eqref{eq:fourth}, finishes the proof of \Cref{thm:main-bi}.

To see why the above claim holds,  we need to present two lemmas.
\begin{lemma}\label{matroidpolytope}
	The point $\{\hat{y}_{i, j}\}\in [0,1]^{n^2}$ is in the matroid polytope, defined as the convex hull of all the independents sets $\mathcal{I}$, corresponding to the laminar matroid $\mathcal{M}=(V,\mathcal{I})$ in \Cref{alg:bicriteria}.
\end{lemma}

\begin{proof}{Proof.}
To prove this, we must show $\sum_{i\in [k]} \sum_{j\in [n]} \hat{y}_{i, j}\leq k,~\forall k\in [n].$ To see why this holds, note the solution $\{\hat{y}_{i,j}\}$ satisfies the first two constraints of \ref{LP1}, and therefore we must have
\begin{align*}
	&\sum_{i\in [k]} \sum_{j\in [n]} \hat{y}_{i, j} \leq \sum_{i\in [k]} \sum_{j\in [n]} \left( \sum_{|S|=i, S\ni j} \hat{x}_{i, S}-\sum_{|S|=i-1, S\ni j} \hat{x}_{i-1, S} \right)\\
	&=\sum_{j\in [n]} \sum_{i\in [k]} \left( \sum_{|S|=i, S\ni j} \hat{x}_{i, S}-\sum_{|S|=i-1, S\ni j} \hat{x}_{i-1, S} \right)\\
	&=\sum_{j\in [n]} \left( \sum_{|S|=k, S\ni j} \hat{x}_{i, S}\right)\leq k \sum_{|S|=k} \hat{x}_{i, S}\leq k.
\end{align*}
Therefore, $\{\hat{y}_{i,j}\}_{i, j\in [n]}$ satisfy the constraints of the matroid polytope of the laminar matroid $\mathcal{M}$. \hfill\Halmos
\end{proof}
The next lemma is the main guarantee of the swap rounding algorithm introduced in \cite{chekuri2010swap}, which is a fundamental property useful for the maximization of monotone submodular functions over matroids.
\begin{lemma}[\cite{chekuri2010swap}]
\label{lem:swap}
Consider a monotone submodular function $f:2^V\rightarrow\mathbb{R}_+$, and a matroid $\mathcal{M}=(V,\mathcal{I})$. Suppose $\vec{y}\in[0,1]^V$ is a feasible point in the matroid polytope $P_{\mathcal{I}}$ of $\mathcal{M}$, and $\hat{S}\in\mathcal{I}$ is the output of the swap rounding algorithm with the input $\vec{y}$. Then:
$$\mathbb{E}\left[ f(\hat{S})\right]\geq \mathbb{E}_{S\sim D^{\textrm{ind}}(\vec{y})}\left[ f(S)\right]~,$$
where $D^\textrm{ind}(\vec{y})$ is the independent distribution with marginals $\vec{y}$, for which the probability of drawing a subset $S\subseteq V$ is
	$\displaystyle\underset{{i\in S}}{\Pi} y_i \underset{{i\notin S}}{\Pi} (1-y_i).$
\end{lemma}
Given \Cref{matroidpolytope} and \Cref{lem:swap}, we have
$
\mathbb{E}\left[G(\hat{\Pi})\right]\geq\mathbb{E}\left[G\left(\Pi^{\textrm{ind}}(\hat{y})\right)\right]
$. Moreover, for each $l\in[L]$ we have 
$
\mathbb{E}\left[G^l(\hat{\Pi})\right]\geq\mathbb{E}\left[G^l\left(\Pi^{\textrm{ind}}(\hat{y})\right)\right]
$, where $\hat{\Pi}\in\mathcal{I}$ is an independent set (in fact base) of the laminar matroid returned by the swap rounding algorithm in \Cref{alg:bicriteria}. The proof of our claim is then immediate as the post-processing step, i.e., returning a permutation $\pi$ over $[n]$ given a base $\hat{\Pi}$ of the laminar matroid $\mathcal{M}$, can only increase the value of sequential submodular functions appearing in the objective and in the group constraints of Problem~\ref{pbm:grp} -- thanks to \Cref{claim3}.\hfill\Halmos

\section{An Improved Approximation for Coverage Choice Model} \label{appCoverage}

In \Cref{revenue}, we studied a rounding algorithm that achieved a bi-criteria $\left((1-1/e)^2, (1-1/e)^2\right)$-approximation for Problem~\ref{pbm:grp}. In this section, we focus on a special case of that problem where the submodular functions of interest are the coverage function. For the practical relevance of this problem, we refer the reader to \cite{ferreira}.
We provide a different LP formulation for this problem and present an oblivious rounding algorithm for this LP, that is, a rounding algorithm that only works with the values of the LP solution and is oblivious to the coverage functions we work with (and subsequently, oblivious to the LP coefficients). This allows us to improve over the result of Problem~\ref{pbm:grp} for coverage choice models and get a $(1-1/e, 1-1/e)$-approximate solution. It also provides a simple $(1-1/e)$-approximation algorithm for Problem~\ref{pbm:seq} under coverage choice models.

\subsection{Model and the LP relaxation}
Consider a special case of the product ranking problem described in \Cref{subsec:application}, where each user of type $u \in \setoftypes$ has a patience level $\patience_u$ and an \emph{interest set} $P_u \subseteq [n]$. For simplicity of notations, we further assume $\mathcal{U}$ is finite. If the products are shown in an order imposed by some permutation $\pi$, the consideration set of a user of type $u$ is the set of the first $\patience_u$ items of the permutation. A user of type $u$ is said to be \emph{engaged} with permutation $\pi$ if and only if $P_u \cap \{\pi_1, \ldots, \pi_{\patience_u} \}\neq \emptyset$. In other words, a user of type $u$ will click on a product if and only if there exists at least one item in their interest set $P_u$ that appears in the first $\theta_u$ positions; that is, 
$$\kappa_u(\{\pi_1,\pi_2,\ldots,\pi_{\theta_u}\})=\mathbb{I}\left\{P_u \cap \{\pi_1, \ldots, \pi_{\patience_u} \}\neq \emptyset\right\}$$
Note the function $\kappa_u(\cdot)$ above is monotone and submodular; therefore, any convex combination of this function is also monotone and submodular. Finally, let $q_u$ denote the proportion of users of type $u$ and let $q^l_u$ denote the proportion of users of type $u$ among users in group $l$. 
%

We start by the IP formulation of the problem. For a given permutation $\pi$, let $x\in \{0, 1\}^{n\times n}$ be the indicator matrix where $x_{i, j}$ corresponds to product $j$ appearing in position $i$. Note a user of type $u$ will engage with the permutation corresponding to the indicator matrix $x$ if and only if $\sum_{i=1}^{\patience_u} \sum_{j\in P_u} x_{i, j}\geq 1$. This allows us to introduce another vector $y\in \{0, 1\}^m$, where $y_u\in\{0,1\}$ is indicating whether a user with type $u$ is engaged with the ranking. Hence, a constraint $\sum_{i=1}^{\patience_u} \sum_{j\in P_u} x_{i, j}\geq y_u$ is established.

Finally, to achieve the linear programming formulation of the model, we relax all the indicator variables to assume values in $[0,1]$, thus leading to the following LP for Problem~\ref{pbm:grp}. Note the LP formulation is the same for Problem~\ref{pbm:seq}, but with no group constraints.
\begin{align*} \label{LP3}\tag{\texttt{Coverage-Choice-Model-LP}}
\max \quad
  &\sum_{u\in \setoftypes} q_u y_u & \nonumber \\
    \text{s.t.} \quad  & \sum_{u \in \setoftypes} q^l_u y_u \geq T_l, & \forall l \in [L] \\
    &\sum_{i=1}^{\patience_u} \sum_{j\in P_u} x_{i, j}\geq y_u, &  \forall u\in \setoftypes\\
    &\sum_{i\in [n]} x_{i, j}=1, & \forall j \in [n] \nonumber \\
    &\sum_{j\in [n]} x_{i, j}=1, & \forall i\in [n] \nonumber \\
    &0\leq y_u\leq 1,~0\leq x_{i, j}\leq 1& \forall u\in \setoftypes,\forall i, j\in [n] \nonumber
\end{align*}

As we discussed, the matrix $x$ imposed by any given permutation $\pi$ and the vector $y$ corresponding to the users attracted by $\pi$ form a feasible (integral) solution of \ref{LP3}. This includes the optimal permutation $\pi^*$, too. Therefore, the optimal value of this LP is an upper bound on the maximum user engagement achievable with respect to the group fairness constraints.

\subsection{Rounding the LP Solution}
At the heart of our algorithm for an improved bi-criteria approximation for Problem~\ref{pbm:grp} under coverage choice models lies a randomized rounding of the optimal solution of \ref{LP3}. Our rounding guarantees that no combination of patience level $\patience_u$ and interest set $P_u$ is far worse than how they were in the fractional solution.  This is formalized in the following proposition.  


\begin{proposition}\label{prop1}
	For any given feasible solution $(x, y)$ of \ref{LP3}, it is possible to construct a randomized integral solution $(\tilde{x}, \tilde{y})$ such that all but the first set of constraints of \ref{LP3} are satisfied, and moreover, $\expect{\tilde{y_u}}\geq (1-1/e)y_u$, for every $u \in \setoftypes$. This can be done in polynomial time.
\end{proposition}

Note any integral solution corresponds to a permutation of products. Therefore, the following theorem is an immediate corollary of ~\Cref{prop1} if we consider the optimal LP solution $(x^*, y^*)$. Note also that the consideration of \Cref{thm2} regarding the randomness of our result still holds.

\begin{theorem}
	The randomized rounding of \Cref{prop1} performed over the optimal solution $(x^*, y^*)$ of \ref{LP3}  provides a $(1-1/e)$-approximation for Problem~\ref{pbm:seq} and a bi-criteria $(1-1/e, 1-1/e)$-approximation for Problem~\ref{pbm:grp} under coverage choice models.
\end{theorem}

\revcolor{In order to prove \Cref{prop1}, we will provide a (dependent) randomized rounding method that achieves $(1-1/e)$ approximation in expectation first. A standard transformation from Las Vegas to Monte Carlo algorithms can then be used to achieve an actual $(1-1/e)$ approximation. We defer the proof of \Cref{prop1} and the details on the randomized rounding in this proof to \Cref{apx:proof-of-rounding-coverage}.}


\section{Conclusion and Open Problems}
\label{sec:conclusion}

In this paper, we introduced the class of sequential submodular maximization problems and studied two applications of such problems in the context of online retailing, namely, the problem of maximizing user engagement with and without group constraints. For maximizing user engagement, we presented an optimal $(1-1/e)$-approximation algorithm. For the same problem with group constraints, we provided a bi-criteria $\left((1-1/e)^2,(1-1/e)^2\right)$-approximation algorithm. For the special case of coverage functions, we provided an optimal bi-criteria $\left((1-1/e),(1-1/e)\right)$-approximation algorithm.

\noindent\textbf{Future Directions.} \revcolor{Our main motivating application in this paper was online retail. However,  identifying other applications for our sequential submodular framework would be interesting. In particular, a possible application of Problem~\ref{pbm:grp} is in \emph{rank aggregation} given a set of options. In a candidate model for this problem, one can consider different subgroups of the society, where each subgroup is a mixture of $n$ different types. Type $i$ only has preferences over the first $i$ options. We assume their valuation function over the first $i$ options is monotone with decreasing marginal values (which is a meaningful assumption in the context of valuation functions). Now the question is how to pick a universal ranking, so that all subgroups' social welfare are maximized at the same time (up to approximations). This simple model is clearly an instance of Problem~~\ref{pbm:grp}. Note that rank aggregation also has applies to the setting when a search engine combines rankings from different sources~\citep{dwork2001rank}. We pose generalizing and studying the applicability of this model, as well as finding other applications for sequential submodular maximization, as future directions.}

\revcolor{In the context of online retail, we focused on maximizing the probability of purchase. This objective function is reasonable in practice where platforms tend to optimize for long-term revenue  and have a strong incentive to create customer loyalty. At the same time, one can consider other objective functions such as maximizing customer surplus (defined as the difference between the user's value for the item and the price) or the total revenue (defined as the expected price paid to the platform). These questions introduce new challenges, mostly due to the fact that these functions are not submodular, as opposed to the probability of purchase function, which is submodular for choice models with substitution property. We leave the question of optimizing these objectives and understanding their trade-offs, similar to our results in \Cref{subsec:revenue}, as open questions.}

Another natural open question arising here would be to improve the approximation ratio for the bi-criteria approximation algorithms of \Cref{sec:bicriteria}. This improvement may be achievable via a more nuanced combination of the modification of our exponential-size LP with \emph{contention resolutions schemes}~\citep{chekuri2014submodular}. Another possible idea (requiring a much stronger result) would be to extend the continuous greedy approach of \cite{calinescu2011maximizing} to incorporate submodular constraints.  As another direction, our particular choice model for the ranking problem is indeed an special case of consider-then-choose choice model studied in \cite{aouad2020assortment}. Extending our approximation algorithms to this general setting under relevant assumptions (e.g., laminarity of considerations sets, similar to ours) is another interesting open problem. Finally, in the vein of \Cref{appCoverage}, one may be able to achieve better results for other submodular functions of interest with specific structures.

\section*{Acknowledgment}
The authors would like to thank Kris Ferreira for her helpful comments and insights on an earlier version of this paper.

\setlength{\bibsep}{0.0pt}
\bibliographystyle{plainnat}
\OneAndAHalfSpacedXI
{\footnotesize
 \bibliography{bibliography}}
 
\ECSwitch
\ECDisclaimer

\renewcommand{\theHsection}{A\arabic{section}}
\renewcommand{\theHsection}{A\arabic{chapter}}



\section{Connections to Product Ranking for Online Retail}
\label{subsec:application}



In our setting, the platform chooses a permutation $\pi$ to rank the products $[n]$ presented to a user $u\in\mathcal{U}$, where $\mathcal{U}$ is the space of user types. The user type $u$ is specified by a pair $(\theta_u, \kappa_{u}(\cdot))$.  He or she considers the first $\theta_u$ products in the list and selects a product with probability $\kappa_u(\{\pi_1, \pi_2 \cdots \pi_{\theta_u}\})$. The platform does not know the exact type of the user but knows the type distribution $\mathcal{D}$, which induces a joint distribution over $(\theta_u, \kappa_{u}(\cdot))$. \revcolor{Note that we allow for arbitrary correlations between $\theta_u$ and $\kappa_u$.}

We assume $\kappa_u(\cdot)$ is a monotone non-decreasing and submodular set function for every $u$. This assumption can be justified in several ways. For example, suppose the value of user $u$ for product $i$ is a random variable $V_{ui}$. After considering a subset $S$, $u$ chooses the item with the highest value from the consideration subset $S$, but only if the value of that item is at least $R$, namely, the value of the outside option. In that case, $\kappa_u(S) =  \mathbb{P}[\max_{i \in S}({V_{ui}}) > R]$ is monotone submodular, even for possibly correlated values of $\{V_{ui}\}$. Furthermore, the choice functions used in the assortment-planning literature such as multinomial logit (MNL), mixed MNL, and their variations, satisfy the substitution property, and hence their corresponding probability of purchase function is monotone submodular (see \cite{KFV-08} for definitions of these models and how they are used in practice.).\footnote{Notably, all of these choices functions with substitution property are special cases of the random utility model.} Importantly, another choice function that is monotone non-decreasing and submodular is the \emph{coverage choice function} introduced in \cite{ferreira}. We study this choice function in \Cref{appCoverage}.

Now, we can show how to cast the problem of maximizing user engagement as a sequential submodular maximization stated in Problem~\ref{pbm:seq}. Recall the goal is to 
 pick ordering $\pi$ to maximize the probability of selection of a product,  $\mathbb{E}_{ u \sim \mathcal{D}}[\kappa_u(\cdot)]$. For each $i\in [n]$, define $f_i(\cdot)\triangleq\mathbb{E}_{ u \sim \mathcal{D}}[\kappa_u(\cdot) | \theta_u=i]$ and $\lambda_i\triangleq\mathbb{P}_{(u)\sim \mathcal{D}}[\theta_u=i]$. The probability that the user makes a purchase is the same as the objective function of Problem~\ref{pbm:seq}, where $f_i(\cdot)$'s are monotone submodular functions and $\lambda_i$'s are non-negative. We can define $f^l_i$'s similarly for Problem~\ref{pbm:grp}, where each group has its own conditional distribution of types. More formally, each group $l \in [L]$ is associated with a subset of types $\mathcal{G}_l\subseteq \mathcal{U}$. For example, one can think of these subsets as various minority groups of users defined by specific sensitive features (e.g., race, gender, or age) in display ads, or shoppers with different behaviours (e.g., frequent shoppers, new shoppers, or shoppers interested in a specific brand) in the online shopping application of product ranking. We highlight that our groups $\mathcal{G}_1,\mathcal{G}_2,\ldots,\mathcal{G}_L$ can be overlapping subsets. Given these subsets, we define $f^l_i(\cdot)\triangleq\mathbb{E}_{ u \sim \mathcal{D}}[\kappa_u(\cdot) | \theta_u=i, u\in\mathcal{G}_l]$ and $\lambda^l_i\triangleq\mathbb{P}_{u\sim \mathcal{D}}[\theta_u=i| u\in\mathcal{G}_l]$ for every $l \in [L]$. Note that in the special case where the groups are non-overlapping and form a portion of $\mathcal{U}$, $f_i$'s are basically convex combinations of $f_i^l$'s for $l\in [L]$.

\section{Analysis of The Greedy Algorithm for Problem~\ref{pbm:seq}} \label{greedyanalysis}
Consider the following greedy algorithm.
\begin{algorithm}[H]
\caption{\label{alg:greedy}Greedy Algorithm}
\begin{algorithmic}[1]
\State \textbf{Input}: $n, \{f_i(\cdot)\}_{i\in [n]}, \{\lambda_i\}_{i\in [n]}$.
\State \textbf{Output}: permutation $\pi$ approximately maximizing $\sum_{i\in [n]} \lambda_i f_i(\{\pi_1, \dots, \pi_i\})$.
\State $\pi \leftarrow ().$
\State \textbf{For} $i=1$ to $n$,
\State\quad\quad $p\leftarrow \argmax_{p\in [n]} \sum_{j=i}^n \lambda_j f_j(\{\pi_1, \dots, \pi_{i-1}, p\})-\sum_{j=i}^n \lambda_j f_j(\{\pi_1, \dots, \pi_{i-1}\})$.
\State\quad\quad $\pi \leftarrow (\pi_1, \dots, \pi_{i-1}, p)$
\State return $\pi$.
\end{algorithmic}
\label{alg1}
\end{algorithm}
\begin{proposition}\label{greedyprop}
The greedy algorithm (\Cref{alg:greedy}) is a $1/2$ approximation algorithm for Problem~\ref{pbm:seq}.
\end{proposition}

\begin{proof}{Proof.}
Suppose the optimal permutation is $(y_1, y_2, \dots, y_n)$. For any partial ordered list of elements  given by the sequence $\pi_1,\pi_2,\ldots,\pi_i\in[n]$, let $F(\pi_1,\pi_2,\ldots,\pi_i)$ denote the mapping that maps this sequence to the value 
$$\sum_{j\in [i-1]} \lambda_j f_j(\{\pi_1, \dots, \pi_{j}\})+\sum_{j=i}^{n} \lambda_j f_j(\{\pi_1, \dots, \pi_{i}\})$$
Based on the greedy selection rule, we have:
	\begin{align}\label{eq1}
		F(\pi_1, \pi_2, \dots, \pi_i)-F(\pi_1, \pi_2,\dots, \pi_{i-1})\geq F(\pi_1, \pi_2, \dots, \pi_{i-1}, y_i)-F(\pi_1, \pi_2,\dots, \pi_{i-1})~.
	\end{align}
Now consider the following sum:
	\begin{align}\label{eq2}
		\sum_{1\leq i\leq n} (F(\pi_1, \dots, \pi_i)-F(\pi_1, \dots, \pi_{i-1}))+\sum_{1\leq i \leq n} (F(\pi_1, \dots, \pi_{i-1}, y_i)-F(\pi_1, \dots, \pi_{i-1}))~,
	\end{align}
We claim the value of \cref{eq2} is at least $F(y_1, \dots, y_n)$. The reason is that each function $f_j$ is monotone submodular and the value added to (\ref{eq2}) by function $f_j$ is:
	\begin{align*}
		~&\lambda_j\sum_{i=1}^j f_j(\{\pi_1, \pi_2, \dots, \pi_i\})-f_j(\{\pi_1, \pi_2, \dots, \pi_{i-1}\})+\lambda_j\sum_{i=1}^j f_j(\{\pi_1, \pi_2,\dots, \pi_{i-1}, y_i\})-f_j(\{\pi_1, \pi_2, \dots, \pi_{i-1}\})\\
		&\geq \lambda_j\sum_{i=1}^j f_j(\{\pi_{[1:i-1]},y_{[1:i-1]},\pi_i\})-f_j(\pi_{[1:i-1]},y_{[1:i-1]})+\lambda_j\sum_{i=1}^j f_j(\pi_{[1:i-1]},y_{[1:i-1]},\pi_i,y_i\})-f_j(\pi_{[1:i-1]},y_{[1:i-1]},\pi_i)\\
		&= \lambda_j\sum_{i=1}^j f_j(\{\pi_{[1:i-1]},y_{[1:i-1]},\pi_i,y_i\})-f_j(\pi_{[1:i-1]},y_{[1:i-1]})\\
		 &= \lambda_j f_j(\{\pi_1, \pi_2, \dots, \pi_j, y_1, y_2, \dots, y_j\})\\
		 &\geq \lambda_j f_j(\{y_1, y_2, \dots, y_j\}),
	\end{align*}
	where the first inequality follows by submodularity and the second inequality follows by monotonicity. Summing above inequalities for  $j=1,\dots,n$ proves the claim. Therefore, by (\ref{eq1}), we have:
	\begin{align*}
		2F(\pi_1, \dots, \pi_n)&=2\sum_{1\leq i\leq n} (F(\pi_1, \pi_2, \dots, \pi_i)-F(\pi_1, \pi_2,\dots, \pi_{i-1}))\\
		&\geq \sum_{1\leq i\leq n} (F(\pi_1, \pi_2, \dots, \pi_i)-F(\pi_1, \pi_2,\dots, \pi_{i-1}))+\sum_{1\leq i \leq n} (F(\pi_1, \dots, \pi_{i-1}, y_i)-F(\pi_1, \dots, \pi_{i-1}))\\
		&\geq F(y_1, y_2, \dots, y_n)~,
	\end{align*}
	which shows the approximation ratio of $1/2$.\hfill\Halmos
	\end{proof}
	We next show this approximation ratio is the best possible for \Cref{alg:greedy}, even in the special case of the product ranking problem described in \Cref{subsec:application}.
\begin{example} Consider an instance of the product ranking problem where we have two products $1$ and $2$, and two users $1$ and $2$ that each appear with probability $\frac{1}{2}$. We assume the selection probability functions of the users $\kappa_1(\cdot)$ and $\kappa_2(\cdot)$ are linear, and the probability of click on product $1$ is $1$ for user 1 and 0 for user 2, and the probability of clicking on product $2$ is 0 for user 1 and $1+\epsilon$ for user 2. In this case, the greedy algorithm will pick the ordering $(2, 1)$, which achieves an expected user engagement of $1+\epsilon$, whereas picking the order $(1, 2)$ would achieve an expected user engagement of $2+\epsilon$.
\end{example}

\section{An Alternative Approach to Problem~\ref{pbm:seq}}\label{sec:appAlternate}

Another natural approach to maximize user engagement would have been to maximize $G(.)$ directly over the space of subsets of $\mathcal{V}$ that correspond to permutations. To approximately find such a set, we can treat a permutation as a perfect matching between products and positions and then, approximately maximizing the multi-linear extension of monotone submodular function $G(.)$ defined as

$$G^{\textrm{MLE}}(\mathbf{x})=\sum_{S\subseteq \mathcal{V}} G(S) \Pi_{i^j \in S} x_i^j \Pi_{i^j \in V\setminus S} (1-x_i^j)$$
over the polytope of perfect matchings characterized by the following constraints:
\begin{align*}
	&\sum_{i\in [n]} x_i^j = 1 &\forall j\in [n],\\
	&\sum_{j\in [n]} x_i^j = 1 &\forall i\in [n]~,
\end{align*}
where $x_i^j$ corresponds to $i^j$ getting (fractionally) picked. One can use the continuous greedy algorithm of \cite{calinescu2011maximizing} to obtain a $(1-1/e)$ approximation to this continuous relaxation of the actual problem. Interestingly, as we show in the following example, we can no longer round a fractional solution $\mathbf{x}$ in the perfect matching polytope to an integral matching without any loss in the value of $G^{\textrm{MLE}}(\mathbf{X})$,  as opposed to the matroid polytope where loss-less rounding methods exist.

\begin{example} Suppose $n=4$ and consider functions $f_1, \cdots, f_4$. In addition, we assume $\lambda_i=1/4$ for all $1\leq i \leq 4$. Let $f_1, f_3$, and $f_4$ always be zero. For $f_2$, we assume a coverage function on $\{a, b\}$ where element $1$ and $3$ cover $a$ and elements $2$ and $4$ cover $b$.

Now consider the following assignment probabilities:
$$
 \mathbf{x}= \begin{bmatrix}
    &0.5~~ &0~~ &0.5~~ &0 \\
    &0 &0.5 &0.5 &0 \\
    &0 &0.5 &0 &0.5 \\
    &0.5 &0 &0 &0.5 
  \end{bmatrix}
$$

where the value in row $i$ and column $j$ corresponds to $x_i^j$. This can be uniquely written as $0.5 M_1+ 0.5 M_2$, where $M_1$ and $M_2$ are two perfect matchings (corresponding to permuations $(1,3,2,4)$ and $(3,2,4,1)$. Our multi-linear extension $G^{\textrm{MLE}}(.)$ achieves a higher value on $0.5 M_1+0.5 M_2$ compared with either $M_1$ or $M_2$, thus proving that we cannot round losslessly in the matching polytope.

However, following the same logic as \Cref{thm2}, because any fractional perfect matching is in the matroid polytope of $M$, we can always losslessly round any fractional perfect matching to an independent set in matroid $M$. For this example, we can, for instance, round $\mathbf{x}$ above to:
$$
 \hat{\Pi}= \begin{bmatrix}
    1 & 0 & 1 & 0 \\
    0 & 1 & 0 & 0 \\
    0 & 0 & 0 & 0 \\
    0 & 0 & 0 & 1 
  \end{bmatrix}~,
$$
which is an independent set of the matroid $M$. This rounding can be achieved through the Pipage rounding of \cite{ageev2004pipage}, as used by \cite{calinescu2011maximizing}. Then, through our post-processing step in \Cref{algEngagement}, this matrix becomes
$$
  \begin{bmatrix}
    1 & 0 & 0 & 0 \\
    0 & 1 & 0 & 0 \\
    0 & 0 & 0 & 1 \\
    0 & 0 & 1 & 0 
  \end{bmatrix}~,
$$
simply because the repetition of the first item is removed and the last item moves one position back. Then, we add the remaining items (third item) at the end. The value of the $G(.)$ on this perfect matching is no less than the initial point.
\end{example}

\section{Extensions: Other Objectives and Beyond Submodularity}
\label{sec:extensions}
\revcolor{
\subsection{Product Ranking for Maximizing Revenue}
\label{subsec:revenue}

In order to define the revenue optimization problem, let $r_j$ be the price of item $j$. Also, let $P_i(j, S)$ denote the probability of an arriving customer with patience level $i$ choosing item $j$ when observing an assortment of items $S$, where $|S| \leq i$. Finally, in this section we consider offering truncated permutations of any length in order to accommodate the possibility of achieving higher revenues by offering a smaller set of items. The revenue generated by offering a truncated permutation of items $\tpi = (\pi_1, \pi_2, \cdots, \pi_k)$ where $k \leq n$ is defined as follows.
\begin{equation}
\label{eq:rev}
\Rev(\tpi) = \sum_{i = 1}^n \lambda_i \left(\sum_{j = 1}^{\min(i, k)} r_{\pi_j} \cdot P_i\left(\pi_j, \{\pi_1, \cdots, \pi_{\min(i, k)}\}\right)\right)
\end{equation}

We make the following standard assumptions regarding the purchase probabilities for all $i\in[n]$. Note that these assumptions hold for all random utility models, e.g., see \cite{desir2015capacity}.
\begin{assumption} \label{asm:monotone}
The probability of purchasing some item (in other words the user engagement) cannot decrease if we offer a larger set of items, i.e., if $S \subseteq T$ then $\sum_{j \in S} P_i(j, S) \leq \sum_{j \in T} P_i(j, T)$.
 \end{assumption}
\begin{assumption} \label{asm:dmr}
The probability of purchasing some item (in other words the user engagement), is submodular, i.e., if $f_i(S)=\sum_{j\in S}P_i(j, S)$, then for all $j\in$ and $S\subset T$ we have:
$$
f_i(S\cup\{j\})-f_i(S)\geq f_i(T\cup\{j\})-f_i(T)
$$
 \end{assumption}
 
The revenue maximization problem is defined as finding a distribution $D_{\tpi}$ over (possibly truncated) permutations that achieves the highest possible expected revenue, i.e.,
\begin{equation}
    \label{pbm:rev}
    \underset{D_\tpi}{\textrm{maximize}}~~ \mathbb{E}_{\tpi \sim D_{\tpi}} \Rev(\tpi)
\end{equation}
Note that similar to Problem~\ref{pbm:seq}, we allow randomized solutions yet the optimal solution of the program is always a deterministic truncated permutation ${\tpi}^*$. Let $r_{\max}$ and $r_{\min}$ respectively denote the highest and lowest price among all the items.

In order to provide an approximate solution to the revenue optimization problem, we partition the items into smaller subsets where the price of items in each subset are within a constant factor of each other. Then, we solve the user engagement/purchase rate problem separately for each subset and find the corresponding permutation over the items in the subset. We then report the permutation that generates the highest revenue. The details can be found in \Cref{alg:rev}.
\begin{algorithm}[ht]
\caption{An Approximation Algorithm for Maximizing Revenue}
\revcolor{
\begin{algorithmic}[1]
\State \textbf{Input}: $n, \{P_i(\cdot, \cdot)\}_{i\in [n]}, \{\lambda_i\}_{i\in [n]}, \{r_j\}_{j\in [n]}$; \textbf{Output}: a (possibly truncated) permutation $\tpi$.
\State Partition items $1, 2, \cdots, n$ into $C = \lfloor\ln(r_{\max}/r_{\min}))\rfloor$ subsets, namely, $S_0, S_1, \cdots,$ and $S_C$, where for any item $j \in S_l$ we have $r_{\min} \cdot e^l \leq r_j < r_{\min} \cdot e^{l + 1}$.
\State \label{step:deffunc} For every $i \in [n]$ and $S \subseteq [n]$ with $|S| \leq i$, define $f_i(S) = \sum_{j \in S} P_i(j, S)$.
\State \label{step:useengagement} For each $0 \leq l \leq C$, let $\tpi^{(l)}$ denote the output of \Cref{algEngagement} for finding an approximately optimal solution for maximizing user engagement according to submodular functions $f_i$ when performed on the ground set $S_l$.
\State Let $\hat{l} = \underset{l}{\arg \max} ~~\Rev(\tpi^{(l)})$. 
\State Return $\tpi^{(\hat{l})}$.
\end{algorithmic}
}
\label{alg:rev}
\end{algorithm}
Note that due to Assumptions~\ref{asm:monotone} and \ref{asm:dmr}, the functions $f_i$ defined in Step~\ref{step:deffunc} of the algorithm can be properly used to feed to the input of \Cref{algEngagement} at Step~\ref{step:useengagement}.

\begin{theorem} \label{thm:rev}
\Cref{alg:rev} provides an $O(\ln(r_{\max}/r_{\min}))$-approximation algorithm for the revenue maximization Problem~(\ref{pbm:rev}).
\end{theorem}

\begin{proof}{Proof of \Cref{thm:rev}.}
Let $\tpi^* = (\pi^*_1, \pi^*_2, \cdots, \pi^*_k)$ denote the (truncated) permutation corresponding to the optimal solution of Problem~(\ref{pbm:rev}). We have
\begin{eqnarray*}
\Rev(\tpi^*) &=& \sum_{i = 1}^n \lambda_i \left(\sum_{j = 1}^{\min(i, k)} r_{\pi^*_j} \cdot P_i\left(\pi^*_j, \{\pi^*_1, \cdots, \pi^*_{\min(i, k)}\}\right)\right) \\ 
&=& \sum_{l = 0}^C \left(\sum_{i = 1}^n \lambda_i \left(\sum_{j = 1}^{\min(i, k)} \mathds{1}(\pi^*_j \in S_l) \cdot r_{\pi^*_j} \cdot P_i\left(\pi^*_j, \{\pi^*_1, \cdots, \pi^*_{\min(i, k)}\}\right)\right)\right) \\
&\leq& \sum_{l = 0}^C \left(\sum_{i = 1}^n \lambda_i \left(\sum_{j = 1}^{\min(i, k)} \mathds{1}(\pi^*_j \in S_l) \cdot r_{\pi^*_j} \cdot P_i\left(\pi^*_j, \{\pi^*_1, \cdots, \pi^*_{\min(i, k)}\} \cap S_l\right)\right)\right).
\end{eqnarray*}
The inequality holds as a result of Assumption~\ref{asm:dmr}. Note that by definition, any item in $S_l$ has a price less than $r_{\min} \cdot e^{l + 1}$. Hence, we derive the following.
\begin{eqnarray*}
    \Rev(\tpi^*) &\leq& \sum_{l = 0}^C r_{\min} \cdot e^{l + 1} \left(\sum_{i = 1}^n \lambda_i \left(\sum_{j = 1}^{\min(i, k)} \mathds{1}(\pi^*_j \in S_l) \cdot  P_i\left(\pi^*_j, \{\pi^*_1, \cdots, \pi^*_{\min(i, k)}\} \cap S_l\right)\right)\right) \\
    &=& \sum_{l = 0}^C r_{\min} \cdot e^{l + 1} \left(\sum_{i = 1}^n \lambda_i f_i(\{\pi^*_1, \cdots, \pi^*_{\min(i, k)}\} \cap S_l) \right)
\end{eqnarray*}

Fix any $0 \leq l \leq C$. Let $\tilde{\pi}^{(l)}$ denote a permutation of the items in $S_l$ starting by the subsequence of the items in $S_l$ that appear in $\pi^*$, and appending the rest of the elements $S_l \backslash \{\pi^*_1, \pi^*_2, \cdots, \pi^*_k\}$ to the end in an arbitrary order. By definition, we have $\{\pi^*_1, \cdots, \pi^*_{\min(i, k)}\} \cap S_l \subseteq \{\tilde{\pi}_1, \cdots, \tilde{\pi}_{\min(i, |S_l|)}\}$ for every $i \in [n]$. Due to the monotonicity of $f_i$'s,  we derive that 
$f_i(\{\pi^*_1, \cdots, \pi^*_{\min(i, k)}\} \cap S_l) \leq f_i(\{\tilde{\pi}_1, \cdots, \tilde{\pi}_{\min(i, |S_l|)}\})$. 

Since $\tpi^{(l)}$ is the output of \Cref{algEngagement} over $S_l$, \Cref{thm2} maintains that it is a $(1-1/e)$-approximation for the user engagement problem. Thus,
\begin{equation*}
    \Rev(\tpi^*) \leq \sum_{l = 0}^C r_{\min} \cdot e^{l + 1} \cdot \left(\frac{e - 1}{e}\right) \cdot \left(\sum_{i = 1}^n \lambda_i f_i(\{\tpi^{(l)}_1, \cdots, \tpi^{(l)}_{\min(i, |S_l|)}\}) \right).
\end{equation*}
However, note that all the elements in $S_l$ have prices of at least $r_{\min} \cdot e^{l}$. Consequently, by plugging in the revenue definition of (\ref{eq:rev}), we get the following.
\begin{eqnarray*}
    \Rev(\tpi^*) &\leq& (e - 1) \cdot \sum_{l = 0}^C r_{\min} \cdot e^{l} \cdot \left(\sum_{i = 1}^n \lambda_i f_i(\{\tpi^{(l)}_1, \cdots, \tpi^{(l)}_{\min(i, |S_l|)}\}) \right) \\
    &\leq& (e - 1) \cdot \sum_{l = 0}^C \Rev(\pi^{(l)}).
\end{eqnarray*}
Hence, the (truncated) permutation $\pi^{(\hat{l})}$ with the highest revenue among $\pi^{(l)}$'s will provide at least a fraction $1/\left((e-1)\cdot C\right)$ of the optimal revenue. Noting that $C = \lfloor \ln(r_{\max}/r_{\min})\rfloor$ completes the proof.\hfill\Halmos
\end{proof}}

\subsection{Discussions on Beyond Submodularity: A Blackbox Approach}
\label{subsec:beyondSM}
\revcolor{
As we will sketch in this subsection, our approach in \Cref{sec:analyzing-main-alg} is flexible to be extended to settings beyond submodularity, as long as certain assumptions remain to be valid in such a generalized setting. Recall the proof of \Cref{thm2}, and in particular the reduction we used in this proof (detailed in \Cref{sec:analyzing-main-alg}, part (III) of the analysis). Suppose we have set functions $f_1,\ldots,f_n$ that are not necessarily submodular. We would like to remark that once we  can identify a certain general property over set functions, which we refer to as ``\emph{Property $(\ast)$}'' in this discussion, so that:
\begin{enumerate}[label=(\roman*)]
    \item This property is satisfied by the function $G(\Pi)=\sum_{i\in[n]}\lambda_i f_i\left(S^{\Pi,i}\right)$, defined over the lifted space $\mathcal{V}$ in our earlier reduction,
    \item Under Property~$(\ast)$ the function $G(\Pi)$ can be approximately maximized over any laminar matroid defined over $\mathcal{V}$, or at least it can only be maximized over the particular laminar matroid $\mathcal{M}$ we use in the proof of \Cref{thm2},
\end{enumerate}
then \emph{exactly the same} approximation ratio, whether a constant term or a super constant term, carries over to the corresponding sequential functions maximization problem, i.e., picking a permutations distributions $D_\pi$ maximizing $\mathbb{E}_{\pi\sim D_\pi}\left[\sum_{i = 1}^n \lambda_i f_i(\{\pi_1, \ldots, \pi_i\})\right]$, as long as $f_i(.)$’s are monotone non-decreasing. This observation simply holds, as the proof of \Cref{claim3} only relies on monotonicity of $f_i$'s and the nested property of laminar sets $[1,\ldots,i]$ for $i=1,\ldots,n$. These two properties guarantee that in the post-processing step, given a base of the laminar matroid $\hat\Pi$, we return a permutation $\hat\pi$ so that $F(\hat\pi)\geq G(\hat{\Pi})$, as desired. Given this observation, we can now ask the following question:
\begin{displayquote}
\emph{Besides submodularity, can we think of other natural candidates as Property~$(\ast)$, so that this properties satisfies the two conditions (i) and (ii) above, and also if $f_i$'s satisfy Property~$(\ast)$ then we can conclude that $G$ also satisfies Property~$(\ast)$ in the lifted space $\mathcal{V}$?} 
\end{displayquote}
Again, by looking at the proof of \Cref{proposition1}, we can make the following informal claim: if Property~$(\ast)$ can be described by a group of inequalities, where in each inequality both sides are linear combination of function values at different sets, e.g., similar to the definition of submodularity, then Property~$(\ast)$ holds for $G$ if it holds for $f_i$'s. To formalize this statement further, we consider the following notion of \emph{approximate submodularity} as an example, defined in \cite{das2018approximate}. A function $f:2^{[n]}\rightarrow \mathbb{R}_{+}$ is $\gamma$-approximate submodular if for every $S,T\subseteq[n],S\cap T=\emptyset$, we have:
$$
\gamma\cdot \left(f(S\cup T)-f(S)\right)\leq \sum_{x\in T}f(S\cup\{x\})-f(S)
$$
Note that it can be easily shown that special case of $\gamma=1$ is equivalent to the original definition of submodularity~\citep{das2018approximate}. Now we have the following proposition, whose proof follows exactly the same lines as in the proof of \Cref{prop1} (omitted for brevity).
\begin{proposition}
\label{prop:apxsm}
If set functions $f_1,\ldots,f_n:2^{[n]}\rightarrow\mathbb{R}_{+}$ are $\gamma$-approximate submodular for $\gamma\in[0,1]$, then $G(\Pi)=\sum_{i\in[n]}\lambda_i f_i\left(S^{\Pi,i}\right)$ is $\gamma$-approximate submodular over $\mathcal{V}$ for all $\lambda_1,\ldots,\lambda_n\in\mathbb{R}_{+}$.
\end{proposition}

Maybe more interestingly, as shown in \cite{das2018approximate}, the classic greedy algorithm for maximizing monotone submodular functions still remains approximately optimal when maximizing $\gamma$-approximate submodular functions. In particular, for maximizing  monotone $\gamma$-approximate submodular functions subject to the cardinality constraint, greedy is a  $1-e^{-\gamma}$ approximation algorithm, and for maximizing $\gamma$-approximate submodular functions subject to general matroid constraints greedy is a $1-2^{-\gamma}$ approximation algorithm~\citep{das2018approximate,ajayi2019approximate}.\footnote{\revcolor{To the best of our knowledge, there is no improved approximation factor in the literature for maximizing $\gamma$-approximate submodular functions subject to a general matroid (e.g., by using an approach based on the continuous greedy algorithm and swap rounding); however we conjecture that such an improvement is possible and pose it as an open problem.}}

Putting all the pieces together, this latter result, together with \Cref{prop:apxsm}, implies that we can obtain a $1-2^{-\gamma}$ approximation for sequential function maximization problem when $f_i$'s are $\gamma$-approximate submodular. We leave the possibility of using the continuous greedy algorithm and swap rounding for maximizing $\gamma$-approximate submodular functions, and hence possibly obtaining an improved approximation factor of $1-e^{-\gamma}$ for this problem as an interesting open question for future work. Such an approximation factor, if exists, will carry over to the sequential function maximization version of the problem as we described above.}

\section{Proof of \texorpdfstring{\Cref{lemma:implement}}{}}
\label{apx:proof-of-implement}
	To see the ``if'' direction, assume a flow of $1$ from the source to the sink exists. We can implement layer $i+1$ as follows. Fix the set of the first $i$ elements is $S\subseteq [n]$. We add element $p\in [n]\setminus S$ in position $i+1$ with probability $\ell(S, S\cup \{p\})/x_{i, S}$ where $\ell(S, S\cup \{p\} )$ is the flow going from $v_{i, S}$ to $v_{i+1, S\cup \{p\}}$. By doing so, the probability of any set $T\subseteq [n]$ of size $i+1$ appearing in the first $i+1$ positions is 
\begin{align*}
&=\sum_{S\subset T:\lvert S\rvert=i}\mathbb{P}\left[\textrm{$S$ appears in the first $i$ positions and element $T\setminus S$ is at position $i+1$}\right]\\
&=\sum_{S\subset T, |S|=i} x_{i,S}\cdot\frac{\ell(S, T)}{x_{i,S}}=\sum_{S\subset T, |S|=i} \ell(S, T)=x_{i+1, T}.
\end{align*}

To see the ``only if'' direction, suppose layer $i+1$ is implementable. Let $\mathbb{P}_{p, S}$ be the probability that the policy implementing layer $i+1$ places element $p\in [n]\setminus S$ at position $i+1$, conditioned on the first $i$ elements being set $S$. We define a feasible flow of $1$ as follows. Let the flow from $s$ to each  node $v_{i, S}$ be $x_{i, S}$. Similarly, let the flow from each node $v_{i+1, T}$ to $t$ be $x_{i+1, T}$. For any node $v_{i, S}$ and any $p\in [n]\setminus S$, let the flow from $v_{i, S}$ to $v_{i+1, S\cup \{p\}}$ be  $x_{i, S}\cdot\mathbb{P}_{p, S}$. Note we have $\sum_{p\in S}\mathbb{P}_{p, S}=1$, and therefore the inflow and outflow for each node $v_{i, S}$ are equal. In addition,  due to implementability, we must have for any $T\subseteq [n]$ with $|T|=i+1$, $\sum_{p\in T} x_{i, S\setminus \{p\}}\cdot\mathbb{P}_{p, T\setminus \{p\}}=x_{i+1, T}$, and therefore the inflow and outflow for each node $v_{i+1,T}$ are equal. So, we have a feasible flow of $1$ from $s$ to $t$. \hfill\Halmos

\section{Optimal Value of  \ref{LP1} Relaxation vs.  Optimal Policy}
\label{sec:appendix-example}
We have seen in \Cref{sec:bicriteria} that linear program~\ref{LP1} is a relaxation to the optimal policy. Below we show by an example that the objective value of \ref{LP1} might in fact be strictly more than the total engagement generated by the optimal policy.

\begin{example}
\label{example:numbers}
Consider a setting where we only have one group of users (we therefore drop the indices corresponding to groups). Let $T=0$, $\lambda_i=1/n$ for all $1\leq i\leq n$. Also suppose all the users share the same choice function; that is, all the functions $f_i$ are identical for all $1\leq i \leq n$.

Suppose we have four products $1, 2, 3, 4$. We define sets and their produced engagement as follows:
\begin{align*}
&\{1, 2, 3, 4\}: (0.74: 19/100, 19/100, 18/100, 18/100)
&\{1, 2, 3\}: (0.58: 20/100, 19/100, 19/100)\\
&\{1, 2, 4\}: (0.57: 19/100, 19/100, 19/100)
&\{1, 3, 4\}: (0.57: 19/100, 19/100, 19/100)\\
&\{2, 3, 4\}: (0.58: 20/100, 19/100, 19/100)
&\{1, 2\}: (0.39: 20/100, 19/100)\\
&\{1, 3\}: (0.39: 20/100, 19/100)
&\{1, 4\}: (0.40: 20/100, 20/100)\\
&\{2, 3\}: (0.39: 20/100, 19/100)
&\{2, 4\}: (0.39: 20/100, 19/100)\\
&\{3, 4\}: (0.39: 20/100, 19/100)
&\{1\}, \{2\}, \{3\}, \{4\}: (0.20: 20/100)
\end{align*}
Each entry above specifies the set presented to the user, the value of the solution, and the probability of each product getting picked by the user. For example, the first entry says if the user looks at all four products, the expected value of the objective function would be 74 and the user would pick the first two products with probability $19/100$ and the last two products with probability $18/100$.
A feasible solution to \ref{LP1} for this problem is
\begin{align*}
	&x_{1, \{1\}}=x_{1, \{2\}}=1/2, &x_{2, \{1, 4\}}=x_{2, \{2, 3\}}=1/2\\
	&x_{3, \{1, 2, 3\}}=x_{3, \{2, 3, 4\}}=1/2, &x_{4, \{1, 2, 3, 4\}}=1
\end{align*}
where $x_{i, S}=0$ for everything else. The expected value of this solution is $191.5/4$, whereas feasible policy can achieve anything better than $191/4$. We should note we have used computer-aided methods to find this example.
\end{example}

\revcolor{
\section{Proof of \texorpdfstring{\Cref{mainLemma}}{}}
\label{sec:anlysis-main-bicriteria}
We begin the proof by recalling the linear program \ref{LP1}:
\begin{align*}\tag{\texttt{Primal}}
\max \quad
  & \sum_{i\in[n]}\sum_{S\subseteq [n], |S|=i} \lambda_i x_{i, S}f_i(S) &   \nonumber \\
 \text{s.t.} \quad  &y_{i, j} = \sum_{|S|=i, S\ni j} x_{i, S}-\sum_{|S|=i-1, S\ni j} x_{i-1, S}, &\forall i, j \in [n] \\
    &\sum_{i\in [n]}\sum_{S\subseteq [n], |S|=i} \lambda_i^l x_{i, S} f_i^l(S)\geq T_l, &  \forall l\in [L]\\
&\sum_{S\subseteq [n], |S|=i} x_{i, S}=  1, &  \forall i\in[n]\\
     &y_{i, j} \geq 0, & \forall i, j\in [n]\nonumber\\
    &x_{i, S} \geq 0,~~~~x_{0, \{\emptyset\}}=0.& \forall i\in [n], S\subseteq [n], |S|=i \nonumber
\end{align*}
and its dual linear program \ref{LP2}:
\begin{align} \tag{\texttt{Dual}}
\min \quad
   &\sum_{i\in [n]} \alpha_i-\sum_{l\in [L]}\gamma_l T_l & \nonumber \\
    \text{s.t.} \quad  &\alpha_i + \sum_{j\in S} \beta_{i+1, j}-\sum_{j\in S} \beta_{i, j}\geq \lambda_i f_i(S)+\sum_{l\in [L]} \gamma_l \lambda_i^l f_i^l(S), &  \forall i\in [n-1], S\subseteq [n], |S|=i  \\
    &\alpha_n -\sum_{j\in [n]} \beta_{n, j}\geq  \lambda_n f_n([n])+\sum_{l\in [L]} \gamma_l \lambda_n^l f_n^l([n]), & \nonumber\\
    &\beta_{i, j} \geq 0~~,~~\gamma_l \geq 0. & \forall i, j\in [n]~~,~~\forall l\in [L] \nonumber
\end{align}
We first show a polynomial-time algorithm that approximately optimizes a linear program slightly different from \ref{LP1}, and then show how the obtained solution of this modified LP relates to the solution of the original linear program, \ref{LP1}, to finish the proof.
 
 Define the linear program \ref{modified-primal} to be the same as \ref{LP1}, with the exception that the third set of constraints is replaced with
\begin{equation*}
\label{modified-primal}\tag{\texttt{Modified-Primal}}
\sum_{S\subseteq [n], |S|=i} x_{i, S} = \frac{e}{e-1} \hspace{0.5 in} \forall i\in [n].
\end{equation*}
Also let \ref{modified-dual} be the dual of \ref{modified-primal}. Therefore, \ref{modified-dual} must be the same as \ref{LP2} but the objective value is replaced with
\begin{equation*}
\label{modified-dual}\tag{\texttt{Modified-Dual}}
    \sum_{i\in [n]} \frac{e}{e-1} \alpha_i-\sum_{l\in [L]}\gamma_l T_l.
\end{equation*}
Remember the form of the feasibility polytope in the linear program \ref{LP2} (\Cref{sec:alg-bicriteria-detail}), which is the same as \ref{modified-dual}. We can separate all of the constraints of the linear program~\ref{modified-dual} in polynomial time except the first one. To get around this issue, we use the following result due to \cite{sviridenko2017optimal}.

\begin{proposition}[\citealp{sviridenko2017optimal}]
\label{sublinear}
		For every $\delta>0$, there exists an algorithm with polynomial running time in $n$ and $\frac{1}{\delta}$ that given a monotone increasing submodular function $g:2^X \rightarrow \mathbb{R}_{\geq 0}$, a linear function $l:2^X \rightarrow \mathbb{R}$, and a matroid $\mathcal{M}$, it produces a set $S\in \mathcal{B}(\mathcal{M})$, satisfying
	$$g(S)+l(S)\geq (1-1/e)g(O)+l(O)+\hat{v}\cdot O(\delta) \hspace{0.5in} \forall O\in \mathcal{B}(M),$$
	where $\hat{v}=\underset{e\in X}{\max}~ g(\{e\})$.
\end{proposition}

To apply this result in our context, for any $i\in [n-1]$, let:
\begin{align*}
	&g(S)= \lambda_i f_i(S)+\sum_{l\in [L]}\gamma_l \lambda_i^l f_i^l(S)\\
	&l(S)=-\alpha_i -\sum_{j\in S} \beta_{i+1, j}+\sum_{j\in S} \beta_{i, j}\\
	&\mathcal{M}=([n], \mathcal{I}=\{S\subseteq [n]~|~|S|\leq i\}),
\end{align*}
where $\mathcal{M}$ is the $i$-uniform matroid. Using \Cref{sublinear} by setting $\delta=\frac{\epsilon}{n}$, we can approximately separate the first set of constraints for any $i$. Now, we run the ellipsoid algorithm to solve the dual, but instead of using an exact separation oracle for all the dual constraints, we use the algorithm in \Cref{sublinear} for the first set of constraints. For details of how to run ellipsoid using separation oracles with (multiplicative and additive) approximate guarantees, we refer the reader to \cite{bubeck2015convex}, Chapter 2. In a nutshell, the algorithms first focuses on the feasibility version of the same optimization problem. It then iteratively gets closer to the approximately optimal solution by restricting the polytope of approximately feasible points. This is done by sending a query call to the approximate separation oracle to identify a new \emph{important constraint} and using the resulting hyper-plane to update to the next  Löwner-John ellipsoid. It finally terminates when the volume of the ellipsoid is small enough. We run the ellipsoid method until termination. Let the solution of \ref{modified-dual} obtained by the ellipsoid method using the approximate separation oracle be $\{\alpha^0_i\}_{i\in [n]}, \{\gamma^0\}_{l \in [L]}, \{\beta^0_{i, j}\}_{i, j\in [n]}$, and the list of important constraints it identifies be $\mathcal{L}$. The obtained dual variables must satisfy the guarantee of \Cref{sublinear} upon termination, i.e.,
\begin{equation}
\label{eq:modified-dual}
    \alpha^0_i + \sum_{j\in S} \beta^0_{i+1, j}-\sum_{j\in S} \beta^0_{i, j}\geq (1-1/e) \left( \lambda_i f_i(S)+\sum_{l \in [L]}\gamma^0_l \lambda_i^l f_i^l(S)\right)-c\cdot\frac{\epsilon}{n}, ~~~~~~~~~  \forall i\in [n-1], S\subseteq [n], |S|=i~,
\end{equation}
where $c$ is the constant in the $O(\epsilon)$ term used in Assumption~\ref{sublinear}. Note that due to Assumption~\ref{assum:bounded}, the submodular function $g(S)$ above is bounded by a constant and hence the additive term in \Cref{sublinear} is of order $O(\epsilon)$. Now we define
\begin{align*}
	&\alpha^1_i=\frac{e}{e-1} (\alpha^0_i+c\cdot\frac{\epsilon}{n}), &\forall i\in [n]\\
	&\gamma_l^1=\gamma_l^0, &\forall l\in [L]\\
	&\beta^1_{i, j}=\frac{e}{e-1} \beta^0_{i, j}, &\forall i, j\in [n].
\end{align*}
Let OPT(\ref{LP2}) denote the value of the objective function of the optimal solution to linear program \ref{LP2} and let $\hat{\text{OPT}}$(\ref{modified-dual}) be the objective value of the solution obtained by the ellipsoid method (when using our approximate separation oracle to solve \ref{modified-dual}). Note that $\{\alpha^1_i\}_{i\in [n]}, \{\gamma^1_l\}_{l \in [L]}, \{\beta^1_{i, j}\}_{i, j\in [n]}$ are feasible in the linear program~\ref{LP2}; therefore, we must have
\begin{align}\label{LPineq1}
\hat{\text{OPT}}(\ref{modified-dual}) =  \sum_{i\in [n]} \frac{e}{e-1}\alpha^0_i-\sum_{l\in [L]}\gamma^0_l T_l =\sum_{i\in [n]} \alpha^1_i-\sum_{l\in [L]}\gamma^1_l T_l -\frac{c\cdot e}{e-1}\cdot \epsilon \geq \text{OPT}(\text{\ref{LP2}})-O(\epsilon).
\end{align}

Note that as we run the ellipsoid method using the approximate separation oracle, as described in Chapter 2 of \cite{bubeck2015convex} and we sketched earlier, we send a query call to the separation oracle at each iteration to identify the current iteration's approximate separating hyperplane (which helps with finding the next iteration's ellipsoid). Therefore, the algorithm essentially \emph{detects} a polynomial size subset of dual constraints throughout its run until termination, one for each iteration, that are important (i.e., ignoring the rest of the constraints does not change the optimal dual solution up to the approximation factor used in the separation oracle -- see \cite{bubeck2015convex} for more details). Now consider the set $\mathcal{L}$ of all the constraints detected by the ellipsoid method  while solving \ref{modified-dual} using our approximate separation oracle. Define \texttt{\textcolor{cornellred}{Restricted-Dual}} to be the same as \ref{modified-dual} but only restricted to these constraints. Similarly, define \texttt{\textcolor{cornellred}{Restricted-Primal}} to be the dual of \texttt{\textcolor{cornellred}{Restricted-Dual}}. Both of these linear programs have polynomial size, since the ellipsoid method detects a polynomial number of constraints while solving \ref{modified-dual}. Note that because of LP duality, we simply have:
\begin{equation}
\label{eq:zero}
    \text{OPT}(\texttt{\textcolor{cornellred}{Restricted-Primal}})=\text{OPT}(\texttt{\textcolor{cornellred}{Restricted-Dual}}).
\end{equation}
Next, we aim to compare $\text{OPT}(\texttt{\textcolor{cornellred}{Restricted-Dual}})$ with  $\hat{\text{OPT}}(\ref{modified-dual})$. Note that the solution found by ellipsoid with the approximate separation oracle used in \Cref{sublinear}, which gives the objective value of $\hat{\text{OPT}}(\ref{modified-dual})$,  is essentially the \emph{optimal solution} of an adapted version of \ref{modified-dual} program where (i) we only keep constraints corresponding to those detected by the ellipsoid algorithm in \ref{modified-dual} and restrict the linear program to only these constraints (exactly as in \texttt{\textcolor{cornellred}{Restricted-Dual}}), and (ii) the RHS of first set of constraints is multiplied by $1-1/e$ and relaxed further by subtracting a $c\cdot\frac{\epsilon}{n}$ term (as in \Cref{eq:modified-dual}); so the optimal solution of \texttt{\textcolor{cornellred}{Restricted-Dual}} forms a feasible solution for this adapted version of \ref{modified-dual}, and hence we should have: 
\begin{equation}
\label{eq:first}
    \text{OPT}(\texttt{\textcolor{cornellred}{Restricted-Dual}}) \geq \hat{\text{OPT}}(\ref{modified-dual}).
\end{equation}
By combining \eqref{LPineq1}, \eqref{eq:zero}, and \eqref{eq:first}, we get 
$$\text{OPT}(\texttt{\textcolor{cornellred}{Restricted-Primal}}) \geq \text{OPT}(\text{\ref{LP2}}) - O(\epsilon)= \text{OPT}(\text{\ref{LP1}}) - O(\epsilon).$$

To obtain our final approximate feasible and approximate optimal solution to \texttt{\textcolor{cornellred}{Primal}}, we first solve \texttt{\textcolor{cornellred}{Restricted-Primal}} optimally in polynomial time in $n$, $L$ and $\frac{1}{\epsilon}$ (which is the size of such a linear program, given the running time of the ellipsoid method). This approach obtains $\{y^*_{i, j}\}_{i, j\in [n]}, \{x^*_{i, S}\}_{i\in [n], S\subseteq [n], |S|=i}$, which achieves an objective value at least equal to $\textrm{OPT}(\ref{LP1})- O(\epsilon)$. However, this solution violates the third constraint of \ref{LP1}; yet, it is actually feasible in the \texttt{\textcolor{cornellred}{Modified-Primal}} and hence: 
$$
\sum_{S\subseteq [n], |S|=i} x^*_{i, S} = \frac{e}{e-1}
$$
 To make it satisfy the third constraint in \texttt{\textcolor{cornellred}{Primal}} and approximately satisfy the second constraint, we simply multiply the solution by $1-1/e$ to obtain $\{\hat{y}_{i, j}\}_{i, j\in [n]}, \{\hat{x}_{i, S}\}_{i\in [n], S\subseteq [n], |S|=i}$, which will lower our objective value by a factor of $1-1/e$ but will make all the constraints satisfied except the second constraint. For the second constraint, for each $l\in [L]$ we have  $\sum_{i\in [n]}\sum_{S\subseteq [n], |S|=i} \lambda_i^l \hat{x}_{i, S} f_i^l(S)\geq (1-1/e)T_l$, which finishes the proof of \Cref{mainLemma}.\hfill\Halmos
}

\section{Proof of \Cref{prop1}}
\label{apx:proof-of-rounding-coverage}
 Suppose we wanted to produce a (potentially infeasible) solution $(\hat{x}, \hat{y})$ that satisfied all the LP constraints except possibly the third constraint, which states that each product appears in one position in the permutation. In this scenario, for each position $i\in [n]$, we could choose exactly one product $j(i)$ at random, according to the probabilities $\{x_{i, j}: j\in[n]\}$. Note this is well defined, because $\sum_{j\in [n]} x_{i, j}=1$ for any given position $i\in [n]$. We let $\hat{x}_{i, j(i)}=1$ and also $\hat{x}_{i, j}=0$ for all $j\neq j(i)$. We repeat this process independently at random for all values of $i\in [n]$. (Note this may violate the second LP constraint because one product may appear in multiple positions. We address this issue later.) Finally, we let $\hat{y}_u=1$ if and only if $\sum_{i\leq \patience_u}\sum_{j\in P_u} \hat{x}_{i, j}\geq 1$.
 
 Note $\mathbb{E}[\hat{y}_u]=\mathbb{P}[\hat{y}_u=1]=1-\mathbb{P}[\hat{y}_u=0].$ However,
$$\mathbb{P}[\hat{y}_u=0]=\mathbb{P}[\sum_{i\leq \patience_u}\sum_{j\in P_u} \hat{x}_{i, j}=0]=\mathbb{P}[{i \leq \patience_u}: \sum_{j\in P_u} \hat{x}_{i, j}=0]=\Pi_{i\leq \patience_u}\mathbb{P}[\sum_{j\in P_u} \hat{x}_{i, j}=0],$$
where the last equality is a result of our rounding process being independent for all $u\in [n]$. Recall that according to our rounding process $\mathbb{P}[\sum_{j\in P_u} \hat{x}_{i, j}=0]=\mathbb{P}[j(i)\notin P_u]=1-\sum_{j\in P_u} x_{i, j}$. Hence,
$$\mathbb{P}[\hat{y}_u=0]=\Pi_{i \leq \patience_u}(1-\sum_{j\in P_u}x_{i, j}).$$

However, due to the feasibility of $(x,y)$ and the second constraint in \ref{LP3}, we have $\sum_{i\leq \patience_u} \sum_{j\in P_u} x_{i, j} \geq y_u$. Therefore, according to the inequality of arithmetic geometric means, we have $\mathbb{P}[\hat{y}_u =0]\leq (1-y_u/\patience_u)^{\patience_u}$. Consequently, due to the fact that $0\leq y_u \leq 1$ and $\patience_u \geq 1$, we have \begin{equation} \label{eq:lowerbound}
    \expect{y_u} = \mathbb{P}[\hat{y}_u=1]\geq 1-(1-y_u/\patience_u)^{\patience_u} \geq (1-1/e)y_u.
\end{equation}


As the last step of the proof, we exchange the solution such that it satisfies the third constraint of \ref{LP3} without compromising the bound in \Cref{eq:lowerbound}. To do so, for the items that appear multiple times according to $\hat{x}$, we only show them at the first position at which they appear. This will possibly leave some positions in the permutation empty. We arbitrarily assign all other items (the ones that have not appeared in any position) to the remaining positions. To make this process rigorous, for each item $j$ let $\textrm{first}(j)$ be the first rank where item $j$ is shown according to $\hat{x}$; that is. $\textrm{first}(j)=\min \{i: \hat{x}_{i, j}=1\}$. If item $j$ is not shown anywhere according to $\hat{x}$, we define $\textrm{first}(j)=0$. Let $U=\{j: \textrm{first}(j)=0\}$ be the set of items not shown. Also, let $V=[n]\setminus \{\textrm{first}(j):j\in [n]\}$ be the set of all positions that offer a previously shown item (i.e., not for the first time). Clearly, $|U|=|V|=n'$ for some $n'\in [n]$. We index the elements of $U$ arbitrarily so that $U=\{j_1, j_2, \ldots, j_{n'}\}$. Similarly, let $V=\{i_1, i_2, \ldots, i_{n'}\}$.

For every $j\notin U$, we define $\tilde{x}_{\textrm{first}(j), j}=1$ and $\tilde{x}_{i, j}=0$ for all $i\neq \textrm{first}(j)$. Also, for every $j_k\in U$ where $k\in [n']$, we define $\tilde{x}_{i_k, j_k}=1$ and $\tilde{x}_{i, j_k}=0$ for all $i\neq i_k$. By construction, $\tilde{x}$ defines an integral matching corresponding to a permutation. We also define $\tilde{y}=\hat{y}$.  Note that for every given $u \in \setoftypes$, we have $\sum_{i \leq \patience_u} \sum_{j\in P_u} \tilde{x}_{i, j}=\sum_{j\in P_u} \mathbb{I}\{{\textrm{first}(j)\leq \patience_u}\}$. Hence, if $\sum_{i\leq \patience_u}\sum_{j\in P_u} \tilde{x}_{i, j}=0$, $\textrm{first}(j)> \patience_u$ for all $j\in P_u$. Therefore $\sum_{i\leq \patience_u}\sum_{j\in P_u} \hat{x}_{i, j}=0$ too. Therefore, if $\hat{y}_u=1$, then $\sum_{i\leq \patience_u}\sum_{j\in P_u} \tilde{x}_{u, j}\geq 1$ and subsequently, $\tilde{y}_u=\hat{y}_u \leq 1 \leq \sum_{i\leq \patience_u} \sum_{j\in P_u} \tilde{x}_{i, j}$. Thus, $(\tilde{x}, \tilde{y})$ is an integral solution that satisfies all (except possibly the first) constraints of \ref{LP3}. Also, because $\tilde{y}=\hat{y}$, by using \Cref{eq:lowerbound} we have $\expect{\tilde{y}_u} \geq (1-1/e)y_u$ for every $u\in \setoftypes$, which completes the proof.\hfill\Halmos

\end{document}